\documentclass[dvipdfmx]{aastex}
\usepackage{lscape}

\slugcomment{(Draft ver.)}
\shorttitle{The GI Simulations with DISPH}
\shortauthors{Hosono et al.}
\renewcommand{\vec}[1]{\mbox{\boldmath $#1$}}
\begin{document}
\title{The Giant Impact Simulations with Density Independent Smoothed Particle Hydrodynamics}

\author{Natsuki \textsc{Hosono}\altaffilmark{1,*}, Takayuki R. \textsc{Saitoh}\altaffilmark{2}, Junichiro \textsc{Makino}\altaffilmark{1,2}, Hidenori \textsc{Genda}\altaffilmark{2}, Shigeru \textsc{Ida}\altaffilmark{2}}
\altaffiltext{1}{RIKEN Advanced Institute for Computational Science, Minatojima-minamimachi, Chuo-ku, Kobe, Hyogo 650-0047, Japan}
\altaffiltext{2}{Earth-Life Science Institute, Tokyo Institute of Technology, Ookayama, Meguro-ku, Tokyo 152-8550, Japan}
\altaffiltext{*}{Tel: +81-(0)78-940-5707 / Fax: +81-(0)78-304-4972}
\email{natsuki.hosono@riken.jp}

\begin{abstract}
At present, the giant impact (GI) is the most widely accepted model for the origin of the Moon.
Most of the numerical simulations of GI have been carried out with the smoothed particle hydrodynamics (SPH) method.
Recently, however, it has been pointed out that standard formulation of SPH (SSPH) has difficulties in the treatment of a contact discontinuity such as a core-mantle boundary and a free surface such as a planetary surface.
This difficulty comes from the assumption of differentiability of density in SSPH.
We have developed an alternative formulation of SPH, density independent SPH (DISPH), which is based on differentiability of pressure instead of density to solve the problem of a contact discontinuity.
In this paper, we report the results of the GI simulations with DISPH and compare them with those obtained with SSPH.
We found that the disk properties, such as mass and angular momentum produced by DISPH is different from that of SSPH.
In general, the disks formed by DISPH are more compact: while formation of a smaller mass moon for low-oblique impacts is expected with DISPH, inhibition of ejection would promote formation of a larger mass moon for high-oblique impacts.
Since only the improvement of core-mantle boundary significantly affects the properties of circumplanetary disks generated by GI and DISPH has not been significantly improved from SSPH for a free surface,
we should be very careful when some conclusions are drawn from the numerical simulations for GI.
And it is necessary to develop the numerical hydrodynamical scheme for GI that can properly treat the free surface as well as the contact discontinuity.
\end{abstract}


\section{Introduction}
The origin of the Moon is one of the most important problems in the planetary science.
The giant impact (GI) hypothesis \citep{HD75, CW76} is currently the most popular, since it can solve difficulties that other models face, such as the current angular momentum of Earth-Moon system and Moon's small core fraction compared to the other rocky planets.
According to the GI hypothesis, at the late stage of the terrestrial planet formation, a Mars-sized protoplanet collided with the proto-Earth and produced a circumplanetary debris disk, from which the Earth's Moon is formed.
To examine whether this scenario really works or not, a number of numerical simulations of collisions between two planetary embryos have been carried out \citep[e.g.,][]{B+86, B+87, B+89, CB91, C97, CA01, C04, NS14}.
Most of them were done by using the smoothed particle hydrodynamics (SPH) method which was a widely used particle-based fluid simulation method developed by \citet{L77} and \citet{GM77}.
Recently, however, it is pointed out that the results of numerical simulation of GI by SPH method should be re-examined from the geochemical point of view.

Recent high precision measurement of isotope ratio revealed that it is not easy for the GI hypothesis to reproduce observed properties of the Moon.
The Moon and the Earth have almost identical isotopic composition for oxygen \citep{W+01}, and isotopic ratios chromium \citep{LS98}, titanium \citep{Z+12}, tungsten \citep{T+07} and silicon \citep{G+07}.
This means that the bulk of the Moon should come from the proto-Earth, unless very efficient mixing occurred for all the isotopic elements \citep{PS07}.
On the other hand, in previous numerical simulations of GI, the disk material comes primarily from the impactor, which is likely to have had the different isotopic compositions from that of the Earth.
To solve this problem, several models have been proposed and studied numerically.
These models have the total angular momentum significantly larger than that of the present Earth-Moon system.
Models with a fast rotating proto-Earth \citep{CS12}, a hit-and-run collision \citep{R+12} and a massive impactor \citep{C12} have been proposed.
Although the excess angular momentum is assumed to be removed by the evection resonance  with the Sun \citep[e.g.,][]{CS12}, it may work only a narrow range of tidal parameters \citep{WT15}.
This means that the Moon was formed by a fortuitous event.

Recently, however, it is pointed out that the results of numerical simulations with the standard formulation of SPH (SSPH) is problematic.
It turned out that SSPH has problems in dealing with a contact discontinuity and a free surface.
It is pointed out that these difficulties result in serious problems, such as the treatment of hydrodynamical instabilities \citep[e.g.,][]{O+03, A+07, V+10, M+12}.
This problem arises from the assumption in SSPH that the local density distribution is differentiable, though in real fluid, the density is not differentiable around the contact discontinuity.
As a result, around the contact discontinuity, the density of the low-density side is overestimated and that of the high-density side is underestimated.
Thus, pressure is also misestimated around the contact discontinuity and an ``unphysical'' repulsive force appears.
This unphysical repulsive force causes a strong surface tension which suppresses the growth of hydrodynamical instabilities.
In the GI simulation, since the core-mantle boundary is a contact discontinuity and the planetary surface also has a density jump, the accurate treatment of the contact discontinuities is very important.

We have developed a novel SPH formulation, density independent SPH (DISPH), to solve the problem for the contact discontinuity \citep[][]{SM13, H+13, H13}.
Instead of the differentiability of the density, DISPH requires the differentiability of the pressure.
As a result, DISPH significantly improves the treatment of the contact discontinuity.
Thereby, GI must also be re-investigated by DISPH.
In this paper, we present results of GI simulations performed with DISPH and compare them with those obtained with SSPH by focusing on the treatment of contact discontinuity and its impacts on the results.
We found that DISPH produced significantly different debris disk, which should lead to different moon forming process.
We concluded that the results of GI is sensitive to the numerical scheme and previous numerical simulations of GI should be re-considered.

It is worth noting that \citet{W+06} reported the results of GI by a three-dimensional grid-base method.
They found that the post impact evolution of the disk is different from that of SSPH.
They pointed out that it is due to the poor description of debris disk.
They suggested that the difference may be due to low-resolution for the debris disk in SPH calculations.
However, it could be rather due to their oversimplified polytropic-like EOS. 
Thus, it is not straightforward to compare their results with SSPH.

\citet{C+13} reported the comparison of the results of GI between adaptive mesh refinement (AMR) and SPH and concluded that the predicted moon mass of two methods are quantitatively quite similar.
Although we notice qualitative differences in disk spatial structures in some of these results (for example, the different clump structure between AMR and SPH in Fig. 4 of \citet{C+13}), our DISPH also predicts similar moon masses for the collision parameters that they tested, as we will mention in section 5.
Comprehensive code-code comparison is needed with grid codes, as well as between DISPH and SSPH.
In this paper, we focus on the latter comparison.

Here we do not insist that the results of GI simulations by DISPH are much closer to realistic phenomenon than by SSPH.
While DISPH has been improved for treatment of a contact boundary, both DISPH and SSPH have a problem to treat free surface, i.e., planetary surface.
We here stress that only the improvement for treatment of a contact boundary significantly affects properties of circumplanetary disks generated by GI.
Therefore, we need to be very careful when some definitive conclusions are drawn from the current numerical simulations for GI.
To clarify details of Moon formation, it is necessary to develop the numerical hydrodynamical scheme for GI that properly treats the planetary surface as well as the core-mantle boundary.

This paper is organized as follows.
In section 2, we briefly describe the numerical technique.
We focus on the implementation of DISPH for non-ideal EOS.
In section 3, we describe models of the GI simulations.
In section 4, we show the results and comparisons of the GI simulations with the two methods and clarify the reason for the difference in the properties of the generated disks.
We also show the results of single component objects, in addition to those of differentiated objects with core-mantle structure.
The former and latter simulations discriminate the differences due to a free surface from a core-mantle boundary between the two methods.
In section 5, we summarize this paper.

\section{Numerical method}
\subsection{Overview of DISPH}
In the SPH method, the evolution of fluid is expressed by the motions of fluid elements that are called SPH particles.
The governing equations are written in the Lagrangian form of hydrodynamic conservation laws.
The equations of motion and energy of the $i$-th particles are written as follows:
\begin{eqnarray}
\frac{d^2 \vec{r}_i}{d t^2} & = & \vec{a}_i^{\rm hydro} + \vec{a}_i^{\rm visc} + \vec{a}_i^{\rm grav}, \label{eq:motion}\\
\frac{d u_i}{d t} & = & \left( \frac{d u_i}{d t} \right)^{\rm hydro} + \left( \frac{d u_i}{d t} \right)^{\rm visc}, \label{eq:energy}
\end{eqnarray}
where $\vec{r}$, $\vec{a}$, $u$ and $t$ are the position vector, the acceleration vector, the specific internal energy and the time, respectively.
The subscript $i$ denotes the value of $i$-th particle.
The superscripts $\rm hydro$, $\rm visc$ and $\rm grav$ mean the contributions of the hydrodynamical force evaluated by SPH, viscosity and self-gravity, respectively.
The formal difference between SSPH and DISPH is in the form of $\vec{a}_i^\mathrm{hydro}$ and $\left(d u_i/d t \right)^{\rm hydro}$.
The other terms in the right hand side of Eqs. (\ref{eq:motion}) and (\ref{eq:energy}) have the same forms for both methods.
Since the equations of SSPH can be found in the previous literature \citep[e.g.,][]{B+86, M92, C04}, here we will only describe those of DISPH.

DISPH is originally developed by \citet{SM13} for the ideal gas EOS and then extended to an arbitrary EOS by \citet{H+13}.
The main advantage of DISPH is the elimination of unphysical surface tension which rises at the contact discontinuity.
The unphysical surface tension in SSPH comes from the requirement of the differentiability of the density.
\citet{SM13} developed a new SPH formulation which does not require the differentiability of the density, but requires that of (a function of) pressure.
As a result, their new SPH can correctly handle hydrodynamical instability.

Note that around the shock, neither the pressure nor the density is continuous.
Thus the assumption of the differentiability of pressure and density is broken across the shock.
\citet{SM13} used the quantity $p^\alpha$ where $p$ is as pressure and $\alpha$ is a constant exponent for their formulation to improve the treatment of the shock\footnote{Note that in \citet{SM13}, they used symbol $\zeta$ instead of the symbol $\alpha$.}.
\citet{SM13} applied $\alpha = 0.1$ and show that the treatment the strong shock is improved.
However, they considered only ideal gas EOS.
Here, we apply this formulation to fluids with non-ideal EOS.
The choice of $\alpha$ is related with how to treat a free surface such as a planetary surface.
DISPH overestimates the pressure gradient around the free surface, while SSPH underestimates it.

In Appendix \ref{Sec:App1}, we show the results of the 3D shock problem simulated with SSPH and DISPH.
In the ideal gas EOS case (Fig. \ref{fig:st3D}), the numerical pressure blip at the contact boundary is the smallest for $\alpha = 0.1$, while the numerical density blip is relatively large for the same $\alpha = 0.1$.
\citet{SM13} determined $\alpha = 0.1$ as a best choice for ideal gas through more detailed discussions.
On the other hand, in the Tillotson EOS case, it is not clear that which $\alpha$ is the best choice.
In the case of 3D shock problem (Fig. \ref{fig:st3D}), $\alpha = 1$ is the best choice.
However, in the case of strong shock test, $\alpha = 0.1$ may be the best choice.
Tillotson EOS (Fig. \ref{fig:st3Dt}) shows the different dependence of the numerical pressure blip at the contact boundary on $\alpha$ from that with ideal EOS.
This indicates that the smaller $\alpha$ works better to treat large pressure jump, namely, the strong shock and the free surface.
We adopt $ \alpha = 0.1$ also in this paper, although more comprehensive tests on the choice of $\alpha$ are needed in future study.
We will show the effects of improved treatment of the free surface and the core-mantle boundary in section 4.1 and 4.2, respectively.

In Appendix \ref{Sec:App2}, we also show the results of the Keplerien disk test and Gresho vortex problem.
Results of the Keplerien disk tests with SSPH and DISPH tell us that both schemes can maintain the disk structure during the first several orbital period, whereas catastrophic breakup takes place before $10$ orbital period.
Previous studies also reported the same results \citep{CD10, H15}.
In our simulations of GI, we only follow about $3.4$ times of the rotation period ($7.0$ hrs at $R = 2.9 R_\mathrm{E}$, where $R_\mathrm{E}$ is the current Earth's radius).
We thus consider that the effect of the numerical AM transfer is not crucial for our simulation results.
From the Gresho vortex test, we can see that there is no critical difference between results with two schemes.
Overall, both SSPH and DISPH are capable of dealing with rotation disks with similar degree, as far as the simulation time is less than $3.4$ orbital periods.

The essential difference between DISPH and SSPH is in the way to estimate the volume element of a particle, $\Delta V_i$.
As the starting point, following \citet{SM13} and \citet{H+13}, we introduce the physical quantity:
\begin{eqnarray}
Y_i = \langle p_i^\alpha \rangle \Delta V_i,
\end{eqnarray}
where brackets mean ``smoothed'' values.
Hereafter, we denotes $p^\alpha$ as $y$.
The value of $y_i$ is given as follows:
\begin{eqnarray}
\langle y_i \rangle & = & \sum_j Y_j W(\vec{r}_i - \vec{r}_j; h_i), \label{eq:smooth_p}
\end{eqnarray}
where $W$ and $h_i$ are the kernel function (see below) and the so-called smoothing length, respectively.
By using $y_i$ and $Y_i$, we derived the equations of motion and energy for our scheme as follows:
\begin{eqnarray}
\vec{a}_i^{\rm hydro} & = & - \sum_j \frac{Y_i Y_j}{m_i} \left[ \frac{\langle y_i \rangle^{1/\alpha - 2}}{\Omega_i} \vec{\nabla} W(\vec{r}_i - \vec{r}_j; h_i) + \frac{\langle y_j \rangle^{1/\alpha - 2}}{\Omega_j} \vec{\nabla} W(\vec{r}_i - \vec{r}_j; h_j) \right], \label{eq:DISPH_motion}\\
\left( \frac{d u_i}{d t} \right)^{\rm hydro} & = & \sum_j \frac{Y_i Y_j}{m_i} \frac{\langle y_i \rangle^{1/\alpha - 2}}{\Omega_i} (\vec{v}_{i} - \vec{v}_{j}) \cdot \vec{\nabla} W(\vec{r}_i - \vec{r}_j; h_i), \label{eq:DISPH_energy}
\end{eqnarray}
where $m_i$ and $\vec{v}_i$ are the mass and the velocity vector of particle $i$, respectively.
Here $\Omega$ is the so-called ``grad-$h$'' term \citep[e.g.,][]{SH02, H13, H+13};
\begin{eqnarray}
\Omega_i = 1 + \frac{h_i}{3 \langle \hat{y_i} \rangle} \frac{\partial \langle \hat{y_i} \rangle}{\partial h_i}.
\end{eqnarray}
Here, $\hat{y}$ is the value to determine the smoothing length;
\begin{eqnarray}
h_i & = & 1.2 \left( \frac{\hat{Y}_i}{\langle \hat{y}_i \rangle} \right)^{1/3},\\
\langle \hat{y}_i \rangle & = & \sum_j \hat{Y}_j W(\vec{r}_i - \vec{r}_j; h_i).
\end{eqnarray}
Note that the choice of $\hat{Y}$ and $\langle \hat{y} \rangle$ is arbitrary, as far as $\hat{Y} / \hat{y}$ has the dimension of volume.
In this paper, following \citet{SM13}, we chose $\hat{Y} = m$ and $\langle \hat{y} \rangle = \langle \rho \rangle$, where $\rho$ is the density.
Note that since the interactions between two particles are antisymmetric, our SPH conserves the total momentum and energy.
The grad-$h$ term improves the treatment of the strong shock \citep[e.g.,][]{SM13, H13}.
In Appendix \ref{Sec:App4}, we show the results of strong shock with DISPH both with and without the grad-$h$ term.
We show that our DISPH with the grad-$h$ term works well for the strong shock.
Our DISPH with the grad-$h$ term has enough capability for the problems which include strong shock.

Note that in order to actually perform numerical integration, we need to determine new values of $\langle y_i \rangle$, by solving a set of implicit equations, Eq. (\ref{eq:smooth_p}) combined with the equation of state $p = p(\rho, u)$.
Thus, as in \citet{H+13}, we iteratively solve Eq. (\ref{eq:smooth_p}).
The number of iterations is set to $3$, following the previous works \citep[e.g., Sec. 5.6 in][]{SM13}.
The iteration procedure is the same as that described in \citet{H+13}, except for the initial guess of $Y_i$.
The initial guess of $Y_i$ is obtained by the numerical integration of $Y_i$ using its time derivative:
\begin{eqnarray}
\frac{d Y_i}{d t} & = & \left( \alpha \gamma_i - 1 \right) \langle y_i \rangle^{1 - 1/\alpha} m_i \frac{d u_i}{d t}\label{eq:DISPH_Y},\\
\gamma_i & = & \frac{\rho_i}{p_i} \left( \frac{\partial p}{\partial \rho}\right)_i^\mathrm{adiabatic}, \nonumber \\
         & = & \frac{m_i}{Y_i} \langle y_i \rangle^{1 - 1/\alpha} \left( \frac{\partial p}{\partial \rho}\right)_i^\mathrm{adiabatic}.
\end{eqnarray}
Note that these equations reduce to those of \citet{H+13} in the case of $\alpha = 1$.

For the kernel function $W$, we employ the cubic spline function \citep{ML85}.
Note that the use of the cubic spline kernel for the derivative sometimes causes the paring instability \citep{DA12, P12}.
In order to avoid this instability, we adopt a gradient of the kernel which has a triangular shape \citep{TC92}.

For the artificial viscosity, with both methods, we use the artificial viscosity described in \citet{M97}.
Note that for both methods we use the smoothed density for the evaluation of artificial viscosity.
The parameter for the strength of the artificial viscosity is set to be $1.0$.
In order to suppress the shear viscosity, we apply the Balsara switch \citep{B95} to the evaluation of the artificial viscosity.

The self-gravity is calculated using the standard BH-tree method \citep{BH86, HK89}.
The multipole expansion is calculated up to the quadrupole order and the multipole acceptance criterion is the same as \citet{BH86}.
The opening angle is set to be 0.75.

\subsection{Timestep}
In the SPH method, the timestep is usually determined by the Courant condition as follows \citep{M97}:
\begin{eqnarray}
\Delta t_i^\mathrm{CFL} & = & C^\mathrm{CFL} \frac{2 h_i}{\max_j v_{ij}^\mathrm{sig}},
\end{eqnarray}
where
\begin{eqnarray}
v_{ij}^\mathrm{sig} = c_i + c_j - 3 \frac{(\vec{v}_i - \vec{v}_j) \cdot (\vec{r}_i - \vec{r}_j)}{|\vec{r}_i - \vec{r}_j|},
\end{eqnarray}
and $c_i$ is the sound speed of particle $i$.
Here $C_\mathrm{CFL}$ is a CFL coefficient which is set to 0.3 in this paper.
In this paper, we consider three additional criteria: fractional changes in the specific internal energy, $Y$ (DISPH only), and the accelerations.
These are
\begin{eqnarray}
\Delta t_i^u & = & C^u \frac{u_i}{|\mathrm{d}u_i/\mathrm{d}t|}, \label{eq:dt_u}\\
\Delta t_i^Y & = & C^Y \frac{Y_i}{|\mathrm{d}Y_i/\mathrm{d}t|}, \label{eq:dt_Y}\\
\Delta t_i^a & = & C^a \sqrt{\frac{h_i}{|\vec{a}_i|}},
\end{eqnarray}
where $C^u, C^Y$ and $C^a$ are dimensionless timestep multipliers.
Throughout this paper, we set $C^u = C^Y = C^a = C^\mathrm{CFL}$.
Note that Eqs. (\ref{eq:dt_u}) and (\ref{eq:dt_Y}) are applied when $d u_i / d t < 0$.

\subsection{Equation of state}
In order to actually evaluate Eqs. (\ref{eq:DISPH_motion}), (\ref{eq:DISPH_energy}) and (\ref{eq:DISPH_Y}), we need the expression of $p(\rho, u)$.
Throughout this paper, we use the Tillotson EOS \citep{T62}, which is widely used in the GI simulations.
The Tillotson EOS contains 10 parameters, which we should choose to describe the given material.
The material parameters of the Tillotson EOS for each material are listed in \citet{M89}, page 234, table AII.3.
Note that in the very low density regime, the Tillotson EOS gives negative pressure which is unphysical on the scale of GI.
To avoid numerical instabilities due to negative pressure, we introduce a minimum pressure $p_\mathrm{min}$ for the Tillotson EOS.
In the scale of GI, the typical value of pressure is order of $\sim 100$ GPa.
Throughout this paper, thus, we set $p_\mathrm{min} = 0.1$ GPa.
Also, we impose the minimum timestep to prevent the timestep from becoming too small due to unphysical values of partial derivatives of EOS by density.
We carefully determine the minimum timestep as one second not to cause poor description of the physical evolution of a system in this paper.
In addition, in this case, we do not evaluate hydrodynamical terms in Eqs. (\ref{eq:motion}) and (\ref{eq:energy}), since this small timestep is actually applied for particles with very low density.

\section{Initial condition}\label{Initial Condition}
We performed numerical simulations of GI from eight initial models.
In this section, we briefly describe how we set up the initial conditions.

We first constructed two initial objects, the proto-Earth (target) and the impactor, which satisfy the given impactor-to-target mass ratio and total mass.
We use $\sim 3 \times 10^5$ SPH particles in total.
Following \citet{B+87}, both objects consist of pure iron cores and granite mantles.
First, we place equal-mass SPH particles in a Cartesian 3D-lattice.
Then, inner $30\%$ of the object is set up as iron and the remaining outer part is set up as granite.
The initial specific internal energy of particles is set to $0.1 G M_\mathrm{E} / R_\mathrm{E}$ J/kg and the initial velocity of particles is set to zero.
Here, $G$ and $M_\mathrm{E}$ are the gravitational constant and the current Earth's mass.
We let the SPH particles relax to the hydrostatic equilibrium by introducing the damping term \citep{M94} to the equation of motion.
The end time of this relaxation process is set to $t = 10000$ seconds, which is about ten times of the dynamical time for the target.
After this relaxation process, the particle velocities for each particle are $\sim 1\%$ of the typical impact velocity (an order of $10$ km/sec in the case of the Moon forming impact).

We constructed eight models.
One of them, model 1.10, corresponds to run \#14 of \citet{CA01}.
They concluded that the Moon would form in this run.
In this model, the impactor approaches the proto-Earth in a parabolic orbit.
Another model, model 1.17, was close to run \#7 of \citet{B+87}.
In this model, the initial relative orbit is hyperbolic.
The remaining models have the same parameters as those of model 1.10 except for the initial angular momentum.
High and low angular momentum models correspond to high-oblique and low-oblique collisions.
In all models, initial objects are non-rotating.
We integrated the evolution of these models for about 1 day.
This duration time of the simulation is smaller than the time scale of numerical angular momentum transfer due to the artificial viscosity \citep[for detail, see][]{C04}.
When we present the result, we set the time of the first contact of two objects as the time zero.

Table \ref{Tb:1} shows the summary of the initial conditions.
The columns indicate the initial separation between two objects ($R_\mathrm{init}/R_\mathrm{E}$), initial angular momentum of the impactor ($L_\mathrm{init}$), velocity at the infinity ($v_\mathrm{\infty}$), the total number of particles ($N_\mathrm{tot}$), the number of particles of target ($N_\mathrm{tar}$), the number of particles of impactor ($N_\mathrm{imp}$), the mass of target ($M_\mathrm{tar} / M_\mathrm{E}$), the mass of impactor ($M_\mathrm{imp} / M_\mathrm{E}$) and the mass of core ($M_\mathrm{core} / M_\mathrm{tot}$).
Here, $L_\mathrm{EM}$ is the angular momentum of the current Earth-Moon system, respectively.
We set $R_\mathrm{E} = 6400$ km, $M_\mathrm{E} = 6.0 \times 10^{24}$ kg and $L_\mathrm{EM} = 3.5 \times 10^{35}$ kg m${}^2$/sec.

\begin{deluxetable}{crrrr}
\tablecolumns{5}
\small
\tablewidth{0pt}
\tablecaption{The model parameters.}
\tablehead{
	\colhead{Parameter}  &
	\colhead{Model 0.88} &
	\colhead{Model 0.99} &
	\colhead{Model 1.05} &
	\colhead{Model 1.10} 
}
\startdata
$R_\mathrm{init} / R_\mathrm{E}$   & 5.0     & 5.0     & 5.0     & 5.0     \nl
$L_\mathrm{init} / L_\mathrm{EM}$  & 0.88    & 0.99    & 1.05    & 1.10    \nl
$v_\mathrm{\infty}$ (km/sec)       & 0       & 0       & 0       & 0       \nl
$N_\mathrm{tot}$                   & 302,364 & 302,364 & 302,364 & 302,364 \nl
$N_\mathrm{tar}$                   & 271,388 & 271,388 & 271,388 & 271,388 \nl
$N_\mathrm{imp}$                   &  30,976 &  30,976 &  30,976 &  30,976 \nl
$M_\mathrm{tar} / M_\mathrm{E}$    & 1.0     & 1.0     & 1.0     & 1.0     \nl
$M_\mathrm{imp} / M_\mathrm{E}$    & 0.109   & 0.109   & 0.109   & 0.109   \nl
$M_\mathrm{core} / M_\mathrm{tot}$ & 0.3     & 0.3     & 0.3     & 0.3     \nl
\enddata
\label{Tb:1}
\end{deluxetable}

\begin{deluxetable}{crrrrrrrr}
\tablecolumns{5}
\small
\tablewidth{0pt}
\tablenum{\ref{Tb:1}}
\tablecaption{Continued.}
\tablehead{
	\colhead{Parameter}  &
	\colhead{Model 1.15} &
	\colhead{Model 1.17} &
	\colhead{Model 1.21} &
	\colhead{Model 1.32}
}
\startdata
$R_\mathrm{init} / R_\mathrm{E}$   & 5.0     & 5.0     & 5.0     & 5.0     \nl
$L_\mathrm{init} / L_\mathrm{EM}$  & 1.15    & 1.17    & 1.21    & 1.32    \nl
$v_\mathrm{\infty}$ (km/sec)       & 0       & 10.0    & 0       & 0       \nl
$N_\mathrm{tot}$                   & 302,364 & 305,389 & 302,364 & 302,364 \nl
$N_\mathrm{tar}$                   & 271,388 & 279,206 & 271,388 & 271,388 \nl
$N_\mathrm{imp}$                   &  30,976 &  26,183 &  30,976 &  30,976 \nl
$M_\mathrm{tar} / M_\mathrm{E}$    & 1.0     & 1.0     & 1.0     & 1.0     \nl
$M_\mathrm{imp} / M_\mathrm{E}$    & 0.109   & 0.1     & 0.109   & 0.109   \nl
$M_\mathrm{core} / M_\mathrm{tot}$ & 0.3     & 0.3     & 0.3     & 0.3     \nl
\enddata
\label{Tb:2}
\end{deluxetable}

\section{Results}
We first show the results of collisions of objects with a single component without core-mantle boundary structure in section \ref{Sec:single comp}.
This set of runs is to discriminate the effects of a free surface from those of core-mantle structure and the core-mantle boundary.
Then we show the results of collisions of two differentiated objects with core-mantle boundary.
We overview the time evolution of eight models obtained with two different methods, DISPH and SSPH in section \ref{Sec:time evolve}.
In all runs, we found the differences between the results with two methods are rather significant.
In section \ref{Sec:Predicted moon mass}, we compare the predicted mass of the moon obtained with two methods.
In section \ref{Sec:SurfaceTension}, we investigate the cause of this difference.

\subsection{Collisions of single-component objects}\label{Sec:single comp}
We consider collisions between single-component planets consisting of only granite mantle.
Here we performed two types of impacts; one is the collision between equal mass objects, the other is the same target-to-impactor mass ratio as described in section \ref{Initial Condition}, but with single-component objects.
Since the objects have no core-mantle boundary, the difference between the results of two methods should come from the treatment of the free surface.

The initial objects are constructed in a similar way to the prescription in section \ref{Initial Condition}, although there is no iron core in this case.
In the run with equal-mass objects, both objects have mass of $1 M_\mathrm{E}$ and radius of $1R_\mathrm{E}$, and the initial specific internal energy is set to be $0.1 G M_\mathrm{E} / R_\mathrm{E}$.
The initial angular momentum is the same as the current angular momentum of the Earth-Moon system.
The velocity of the impactor at infinity is zero.
In this simulation we employ 300,754 particles in total.

\begin{figure}
\plotone{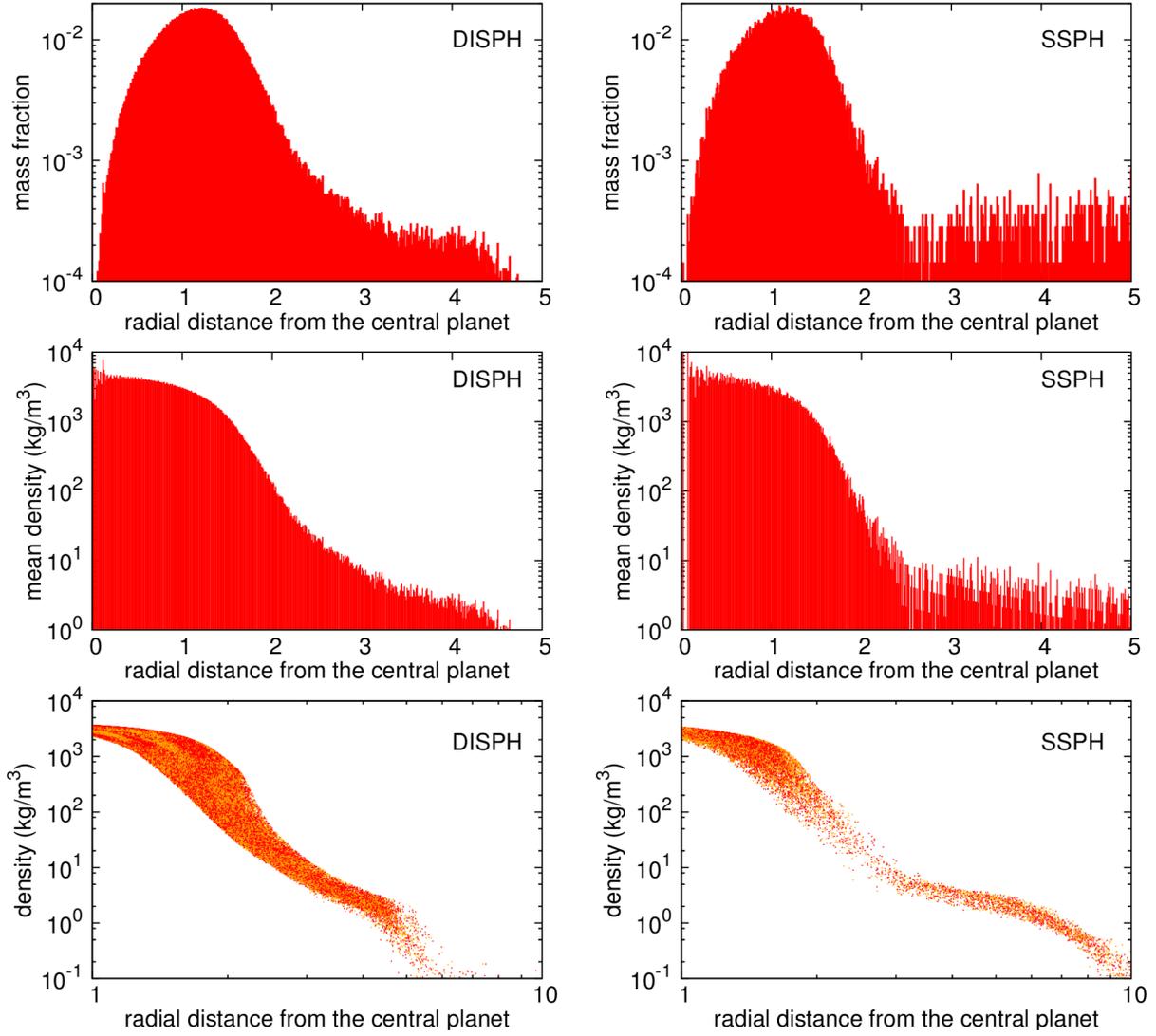}
\caption{
	Radial profiles of the mass and the density for the SSPH and DISPH methods.
	The upper two panels show binned mass in log scale.
	The binned width is set to $0.01 R_\mathrm{E}$.
	The mass is normalized in the current Earth mass and the distance is normalized by current Earth radius.
	The middle two panels show mean mass within the binned volume.
	The lower two panels show radial distance from central planet vs. density for each particle.
	Each axis is shown in log scale.
	The left column shows the results for DISPH and the right column shows those of SSPH.
}
\label{fig:single_comp}
\end{figure}

\begin{figure}
\plotone{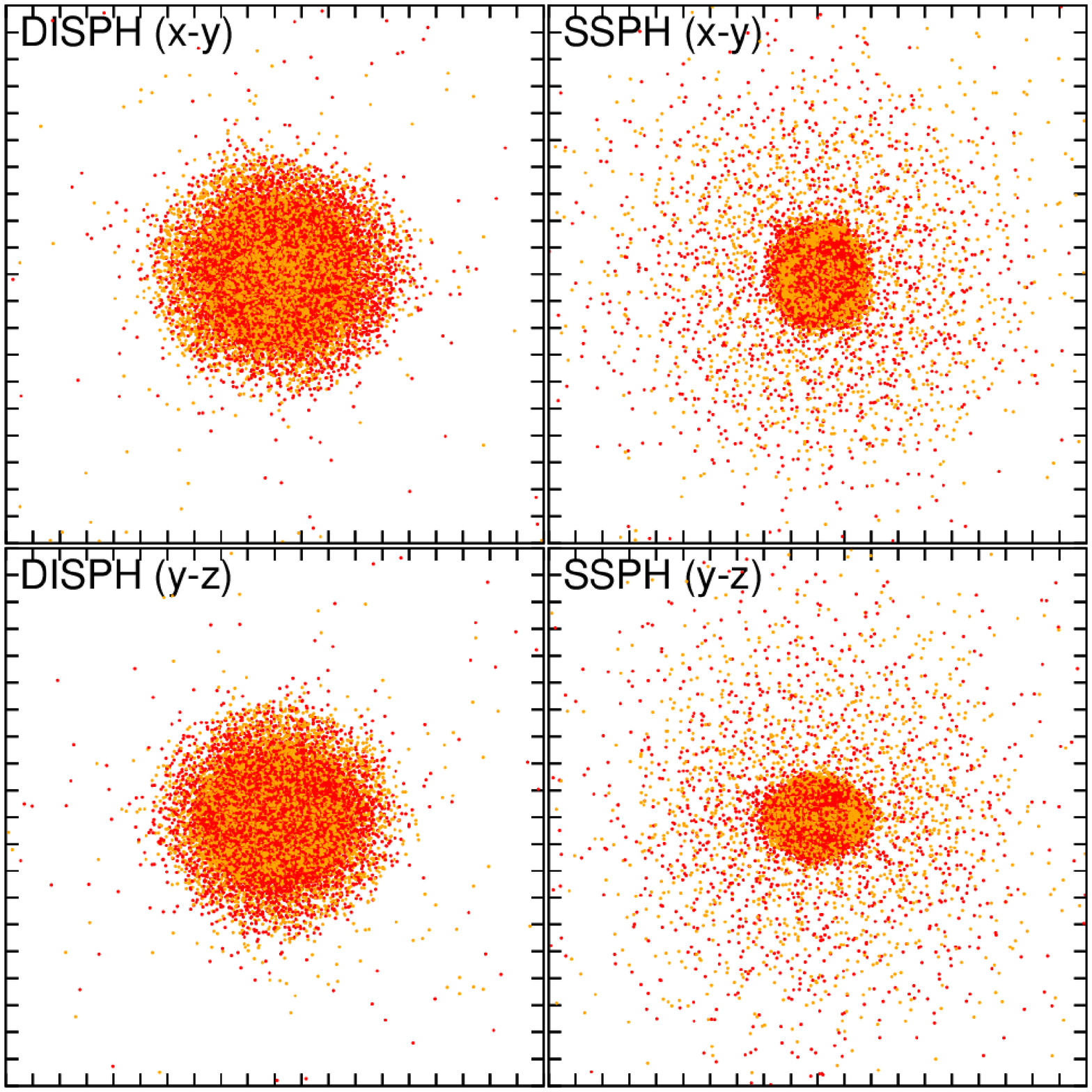}
\caption{
	Snapshots for the collision between the same mass objects which consist of granite.
	The left column shows the results of DISPH and right column shows those of SSPH.
	The upper row is shown in $x-y$ plane and the lower row is shown in $y-z$ plane.
	The color of each particle indicates the original objects of particle.
}
\label{fig:single_comp_snapshots}
\end{figure}

Figure \ref{fig:single_comp} shows the radial profiles of mass and density of the final outcome of the collision of two equal mass objects for both methods.
With SSPH, a gap in the particle distribution is formed around $r \simeq 2.5 R_\mathrm{E}$, while with DISPH the radial distribution is continuous.
This means that SSPH produces gap structure between the body and disk and more spreading disk than DISPH.
The gap at $r \simeq 2.5 R_\mathrm{E}$ is also found in the snapshot on the $x$-$y$ plane with SSPH (Fig. \ref{fig:single_comp_snapshots}).
There is, however, no physical reason for the formation of this gap.
It seems to be natural that the angular momentum distribution is continuous.
Why the gap is formed in the SSPH simulation is most likely the same as the gap formation at the contact discontinuity (for detail, see section \ref{Sec:SurfaceTension}).
Since there is a density jump around the free surface, the free surface is a kind of contact discontinuity.
Though there is no discontinuity in the density distribution, the slope is steep at 2-3 Earth radius and the density itself is low.
Thus, the density difference between two particles radially separated can be very large, resulting in the problem similar to that in the contact discontinuity.
DISPH does not suffer from such a problem.

The result in Appendix \ref{Sec:App3} also suggests that DISPH is better than SSPH for the treatment of the free surface.
However, since around the free surface the pressure is not continuous as well as density, it cannot be readily concluded that DISPH is sufficiently improved from SSPH for treatment of the free surface.

\begin{figure}
\plotone{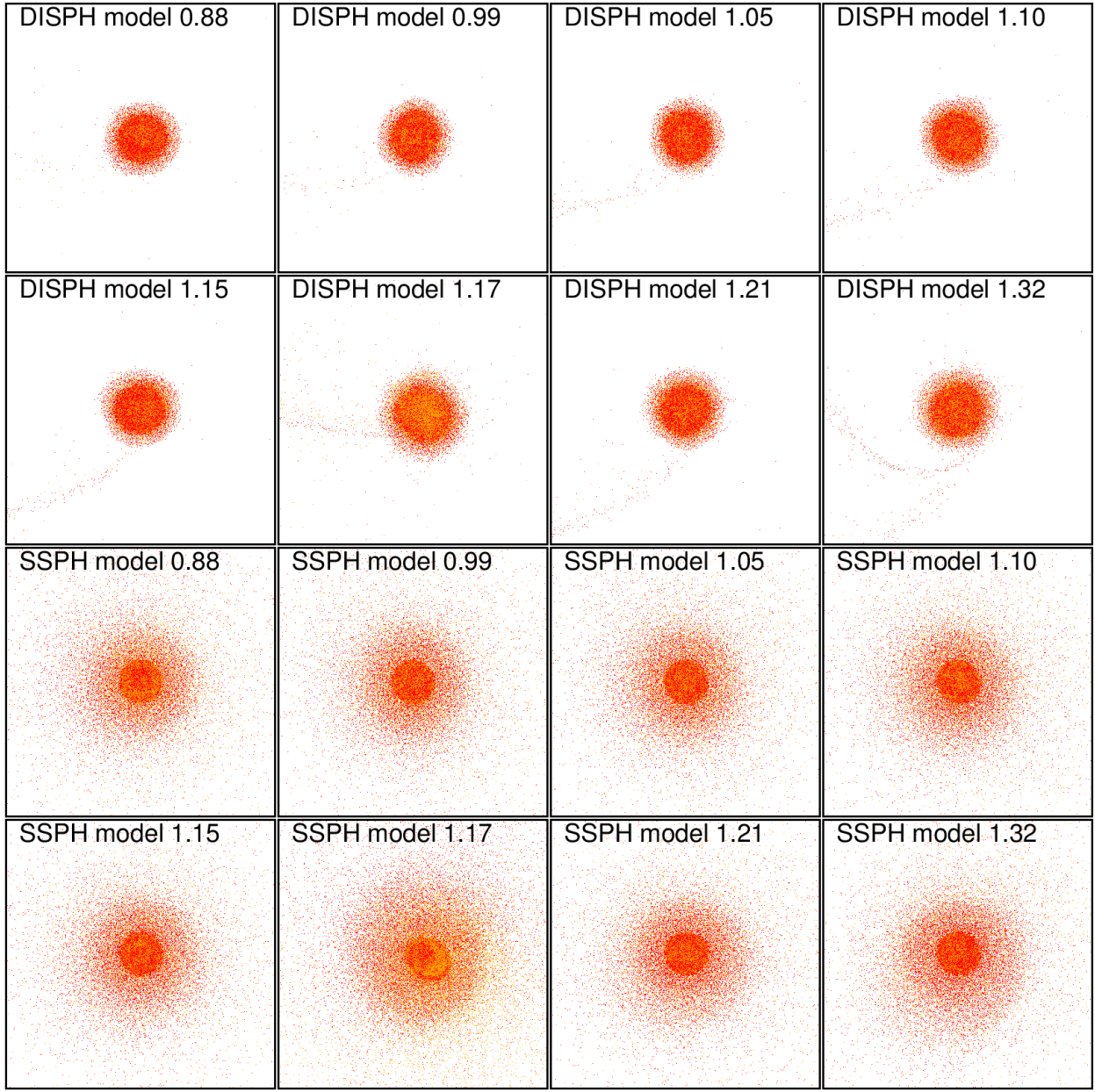}
\caption{
	Snapshots for the collision between single component objects at $t = 24$ hrs.
	The upper two rows show the results of DISPH, while lower two rows show those of SSPH.
	The orange particles indicate material of the target, while red particles indicate those of impactor.
}
\label{fig:single_comp_snapshots2}
\end{figure}

Figure \ref{fig:single_comp_snapshots2} shows the snapshots for a set of runs of single-component objects with the mass ratio of 10:1 given by Table 1.
They show a similar trend on difference between the two methods to that found in Fig. \ref{fig:single_comp}.
DISPH generally tends to produce more compact disks than SSPH does.
Note that the sizes of the planet after an impact are roughly the same between SSPH and DISPH.
Because DISPH produces more compact disks, the results by DISPH look as if the planet itself is inflated.

In order to compare the results between two methods quantitively, we employ the so-called ``predicted moon mass'' as just a reference value.
First, we extract ``disk particles'' from the simulation results.
Following \citet{CA01}, we classified SPH particles to three categories, namely, escaping particles, disk particles and planet particles.
Particles whose total (potential + kinetic) energies are positive are regarded as escaping.
If the total energy of a particle is negative and its angular momentum is greater than that of the circular orbit at the surface of the planet, it is categorized as a disk particle.
Then, other particles are categorized as planet particles, since these particles should fall back to the target.
After the particles are classified, we predict the moon mass using information of disk particles.
According to the $N$-body simulations by \citet{I+97} and \citet{K+00}, the predicted moon mass, $M_\mathrm{M}$, is given by:
\begin{eqnarray}
M_\mathrm{M} = 1.9 \frac{L_\mathrm{disk}}{\sqrt{G M_\mathrm{E} R_\mathrm{Roche}}} - 1.1 M_\mathrm{disk} - 1.9 M_\mathrm{disk,escape}, \label{eq:predMoonmass}
\end{eqnarray}
where $L_\mathrm{disk}$, $R_\mathrm{Roche}$ and $M_\mathrm{disk,escape}$ are the angular momentum of the disk, the Roche radius of the planet, the mass of the disk, respectively.
Note that $M_\mathrm{disk,escape}$ is the total mass of disk particles that escape from the disk through scattering by accreting bodies.
Following previous works \citep[e.g.,][]{K+00, C04}, we set $M_\mathrm{disk,escape}$ to $0.05 M_\mathrm{disk}$.
Assuming that materials from the proto-Earth and the impactor are well mixed in the disk particles, we estimate the fraction of the moon materials originating from the proto-Earth.
It has been known that the Moon and the Earth have identical isotope ratios for several elements.
This means that the Moon should contain large fraction ($> 90 \%$) of materials from proto-Earth mantle \citep[e.g.,][]{C12, CS12, R+12}.

Note that since Eq. (\ref{eq:predMoonmass}) is an empirical equation, this equation sometimes yields an unphysical moon mass, such as a negative mass or a greater mass than the disk, in particular for high-oblique impacts.
However, $M_\mathrm{M}$ in Eq. (\ref{eq:predMoonmass}) is a good indicator for quantitative comparison between DISPH and SSPH.
In this paper we use $M_\mathrm{M}$ in Eq. (\ref{eq:predMoonmass}) as a reference value for the comparison.

\begin{figure}
\plotone{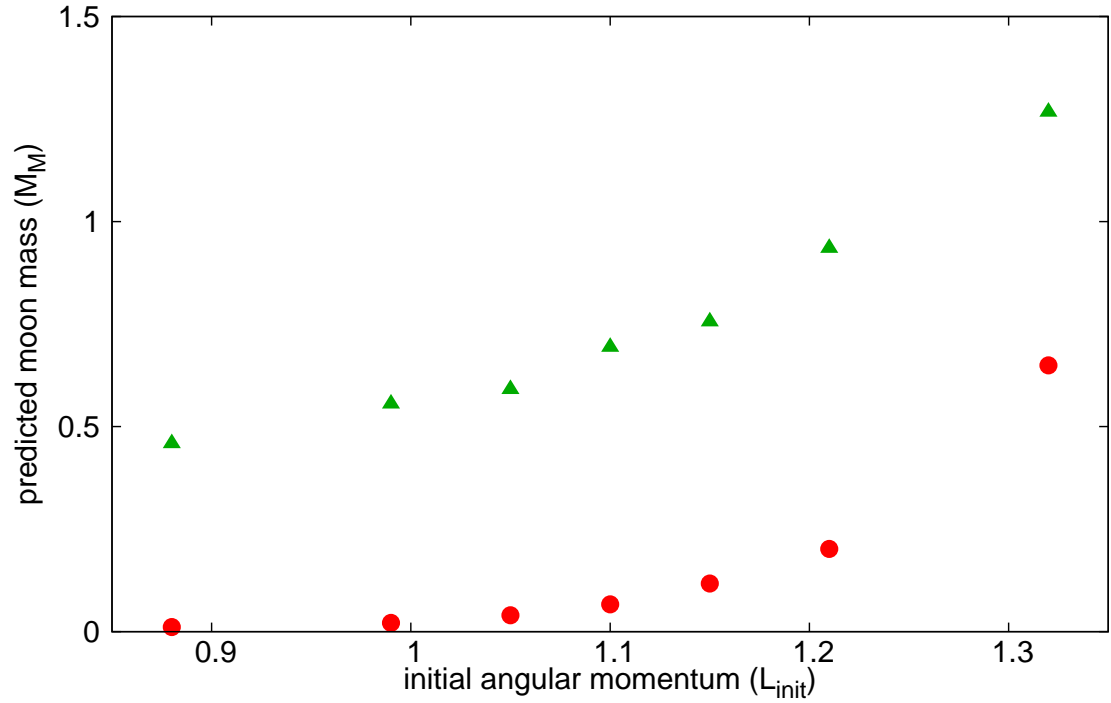}
\caption{
	The initial angular momentum vs. predicted moon mass at $t = 24$ hrs with both methods.
	Red circles indicate the results with DISPH, while green triangles indicate those of SSPH.
	The angular momentum is normalized in the current angular momentum of Earth-Moon system.
	The predicted moon mass is normalized in the current Moon mass.
	Note that model 1.17 is not plotted in this figure.
}
\label{fig:single_comp_AMvsMM}
\end{figure}

Figure \ref{fig:single_comp_AMvsMM} shows the predicted moon mass ($M_\mathrm{M}$) as a function of initial impact angular momentum ($L_\mathrm{init}$), obtained by the collisions of two single-component objects.
Both methods have a qualitatively similar $L_\mathrm{init}$-dependence; $M_\mathrm{M}$ increases as $L_\mathrm{init}$ increases.
However, DISPH produces more compact disks and accordingly smaller $M_\mathrm{M}$ than SSPH does.
Since the two objects have no core-mantle boundary, this difference should come from the treatment of the free surface.

\begin{figure}
\plotone{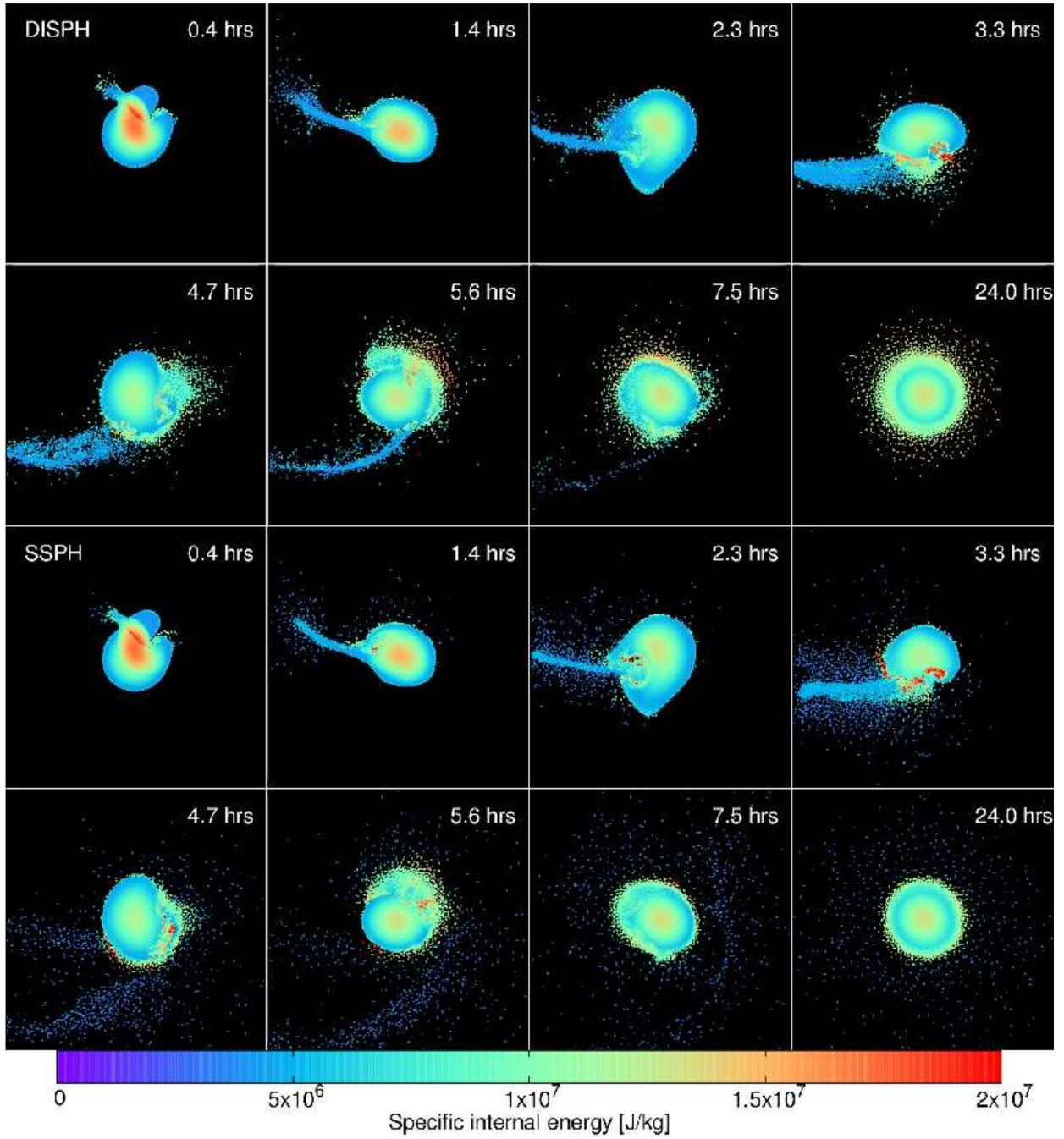}
\caption{
	Specific internal energies for model 1.15 within $0.1 R_\mathrm{E}$ slice with both methods are shown.
	The color bar is given at the bottom.
	Top two rows show the results of DISPH, while bottom two rows show those of SSPH.
}
\label{fig:temp}
\end{figure}

Figure \ref{fig:temp} shows the distributions of specific internal energy for model 1.15 with both methods.
The difference between two methods are clear.
The first two snapshots for each method look fairly similar; shock heating and the arm-like structure can be clearly shown.
In the panels $t > 2.3$ hrs, however, clear difference between two methods can be seen.
With DISPH, the arm re-collides to the proto-Earth and undergoes shock heating again, which results in hot and compact debris disk (panel $t = 24.0$ hrs).
On the other hand, with SSPH, cold particles are ejected around the arm-like structure (panels $t = 2.3-3.3$ hrs).
These ejected particles finally become the cold and expanded disk (panels $t \geq 4.7$ hrs).
Note that similar cold and expanded disk can be seen in previous studies with SSPH.
This difference might come from the treatment of free surface or shock.
Since in this paper we focused on the treatment of core-mantle boundary, further investigation of the origin of this difference is left for future works.

We will show a similar plot for collisions between differentiated objects.
As we will show in Fig. \ref{fig:AMvsMM}, the dependance of $M_\mathrm{M}$ on $L_\mathrm{init}$ is different from that in Fig. \ref{fig:single_comp_AMvsMM}.
The difference is due to the contribution from core-mantle structure and its boundary.
The core-mantle boundary has to be treated properly as well as the free surface in GI simulations.

\subsection{Collision of objects with core-mantle structure}\label{Sec:time evolve}
\begin{figure}
\plotone{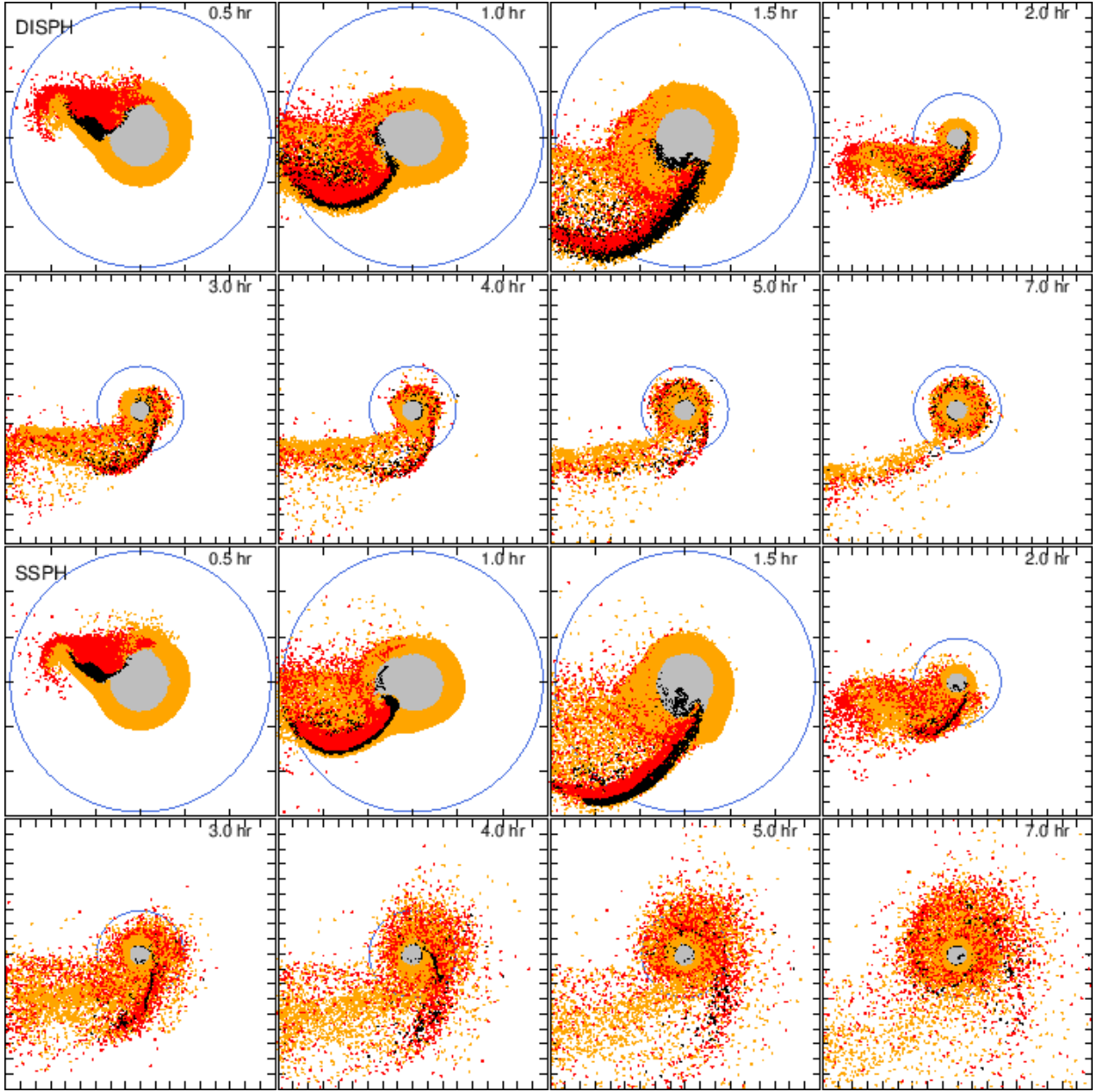}
\caption{
	The upper two rows show time series of the model 1.10 with DISPH, while the lower two rows show those with SSPH.
	Results are shown in the face on view and particle with $z \leq 0$ are shown.
	Times are $t = 0.5, 1.0, 1.5, 2.0, 3.0, 4.0, 5.0$ and $7.0$ hrs from the initial contact of two objects.
	The unit length is set to the current radius of Earth, $R_\mathrm{E}$.
	In the first three panels, the length of each side is $6 R_\mathrm{E}$ and in the other panels the length of each side is $20 R_\mathrm{E}$.
	The orange and gray particles are mantle and core particles of the target and red and black particles are those of the impactor.
	The blue circles indicate the Roche limit, $\sim 2.9 R_\mathrm{E}$.
}
\label{fig:RUN1_X}
\end{figure}

Next we discuss details of collisions between differentiated objects with core-mantle structure.
We will show that the contact density discontinuity at the core-mantle boundary may be a problem in calculations with SSPH and it is improved with DISPH.
Figure \ref{fig:RUN1_X} shows the time evolution of model 1.10 obtained with two methods.
What is shown here is the face-on view (in the $x$-$y$ plane) of particles with negative values of the $z$ coordinate, seen from $z = \infty$.

The two runs in the first frame ($t = 0.5$ hr) look fairly similar.
In both runs, the core of the impactor (black) is significantly deformed.
It pushes up the mantle of the proto-Earth (orange) and the mantle of the impactor (red) is left behind.
In the first four frames, the results of the two models are qualitatively similar.
In both runs, the core of the impactor forms an arc-like structure at $t = 1$ hr, which becomes more extended at $t = 1.5$ hr.
However, this arc is more extended for the run with SSPH.
In the frames of $t = 2$ hr through $t = 7$ hr, most of the mass of the impactor, which has not escaped from the planet gravitational potential, fall back to the proto-Earth in the run with DISPH, while some of the mass forms extended envelope and disk in the run with SSPH.
As we will see in the next section, this difference in the structure causes a large difference in the predicted moon mass.

\begin{figure}
\plotone{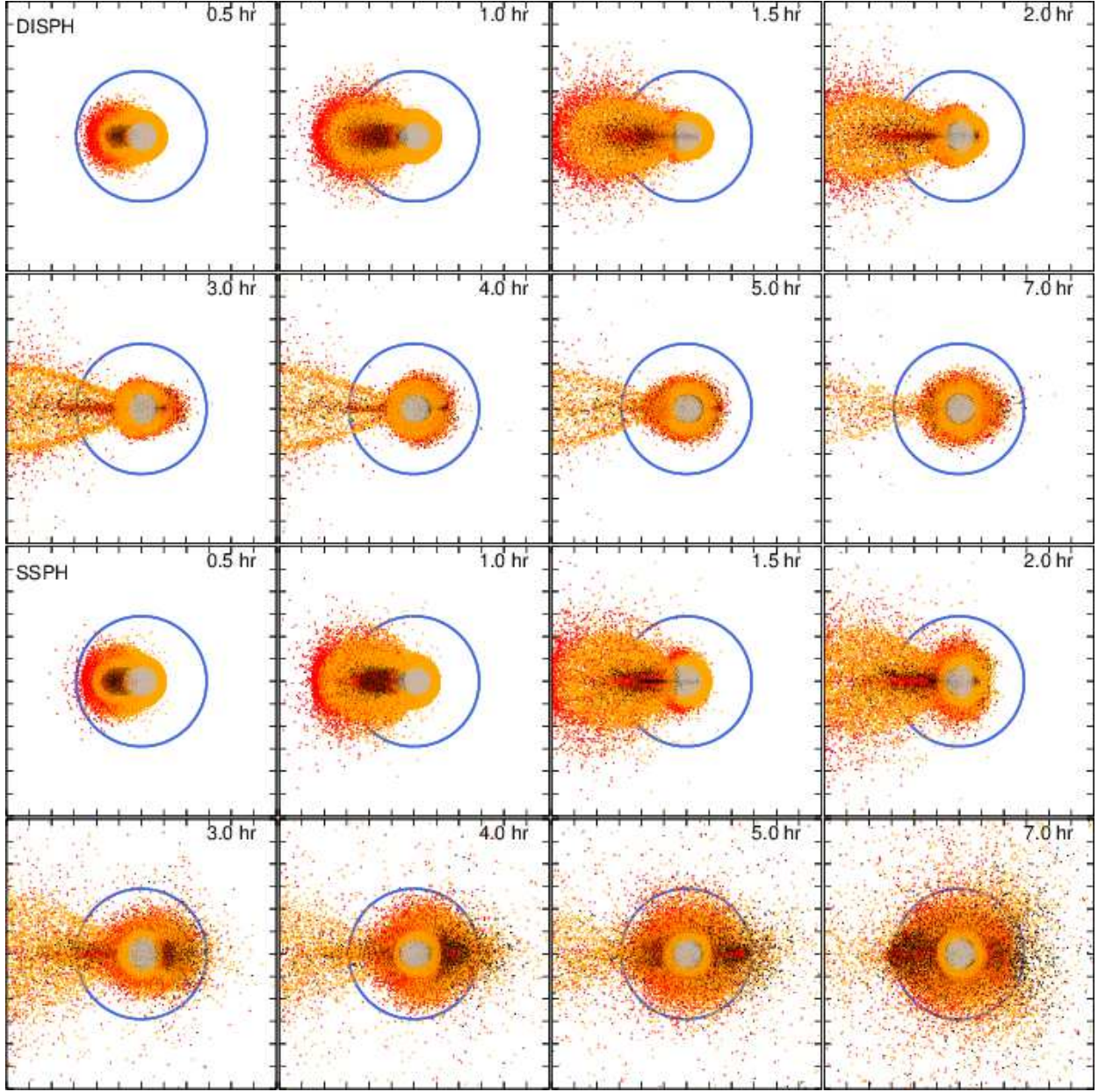}
\caption{
	The edge-on views of results in Fig. \ref{fig:RUN1_X}.
	The length of each side is $12 R_\mathrm{E}$.
}
\label{fig:RUN1_OE}
\end{figure}

Figure \ref{fig:RUN1_OE} shows the edge-on views of the two runs of model 1.10.
The vertical distribution of ejected mantle material is also quite different.
In the frames of $t = 1$ hr and $2$ hr, the results are qualitatively similar.
In the frames of $t = 3$ hr through $7$ hr, however, SSPH produces vertically stretched disk.
On the other hand, with DISPH, the number of the disk particles is much smaller than with SSPH.
DISPH produces a thinner disk than SSPH.

\begin{figure}
\plotone{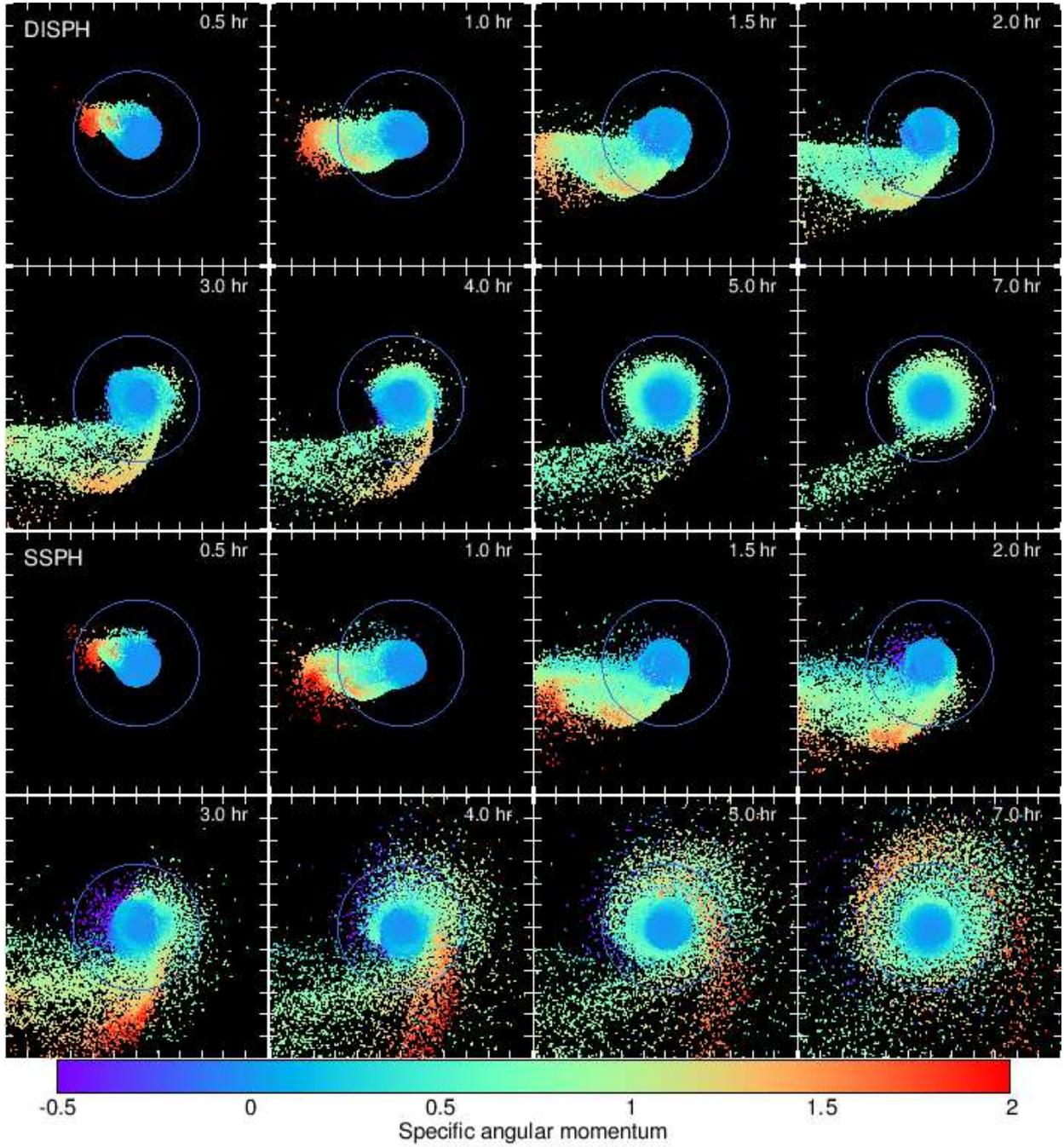}\\
\caption{
	Same as Fig. \ref{fig:RUN1_X}, but the contour is specific angular momentum around $z$-axis is shown.
	The length of each side is $12 R_\mathrm{E}$.
	The color scale is normalized by $\sqrt{G M_\mathrm{E} R_\mathrm{E}}$.
}
\label{fig:RUN1_Ang}
\end{figure}

Figure \ref{fig:RUN1_Ang} shows the distribution of the angular momentum.
For $t \ga 2$ hr, the ejecta with high angular momentum ($> \sqrt{GM_\mathrm{E}R_\mathrm{E}}$) is much abundant in the SSPH result (lower panels) than in the DISPH result.

\begin{figure}
\plotone{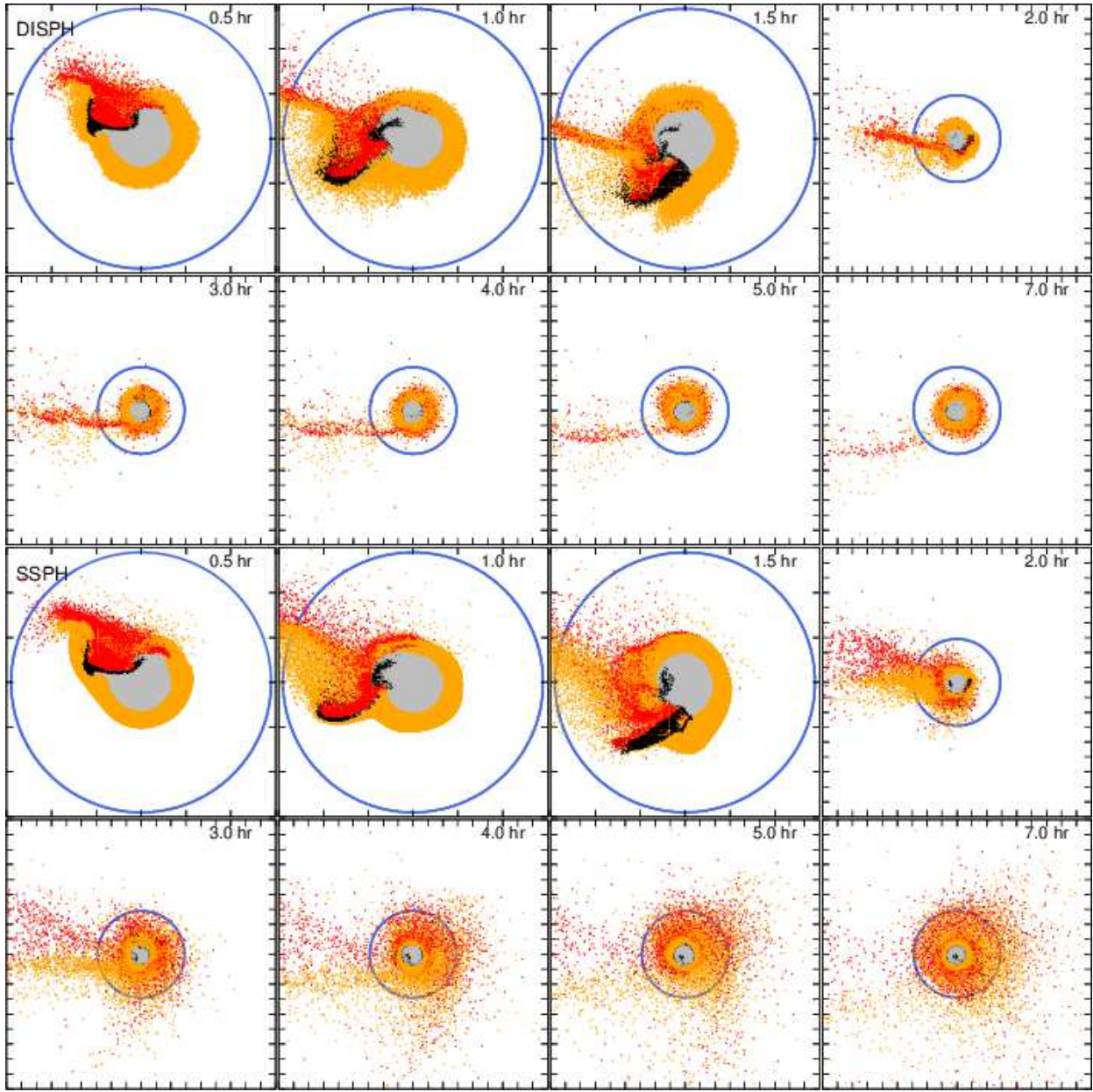}
\caption{
	Same as Fig. \ref {fig:RUN1_X}, but for model 0.88.
}
\label{fig:RUN4_X}
\end{figure}

\begin{figure}
\plotone{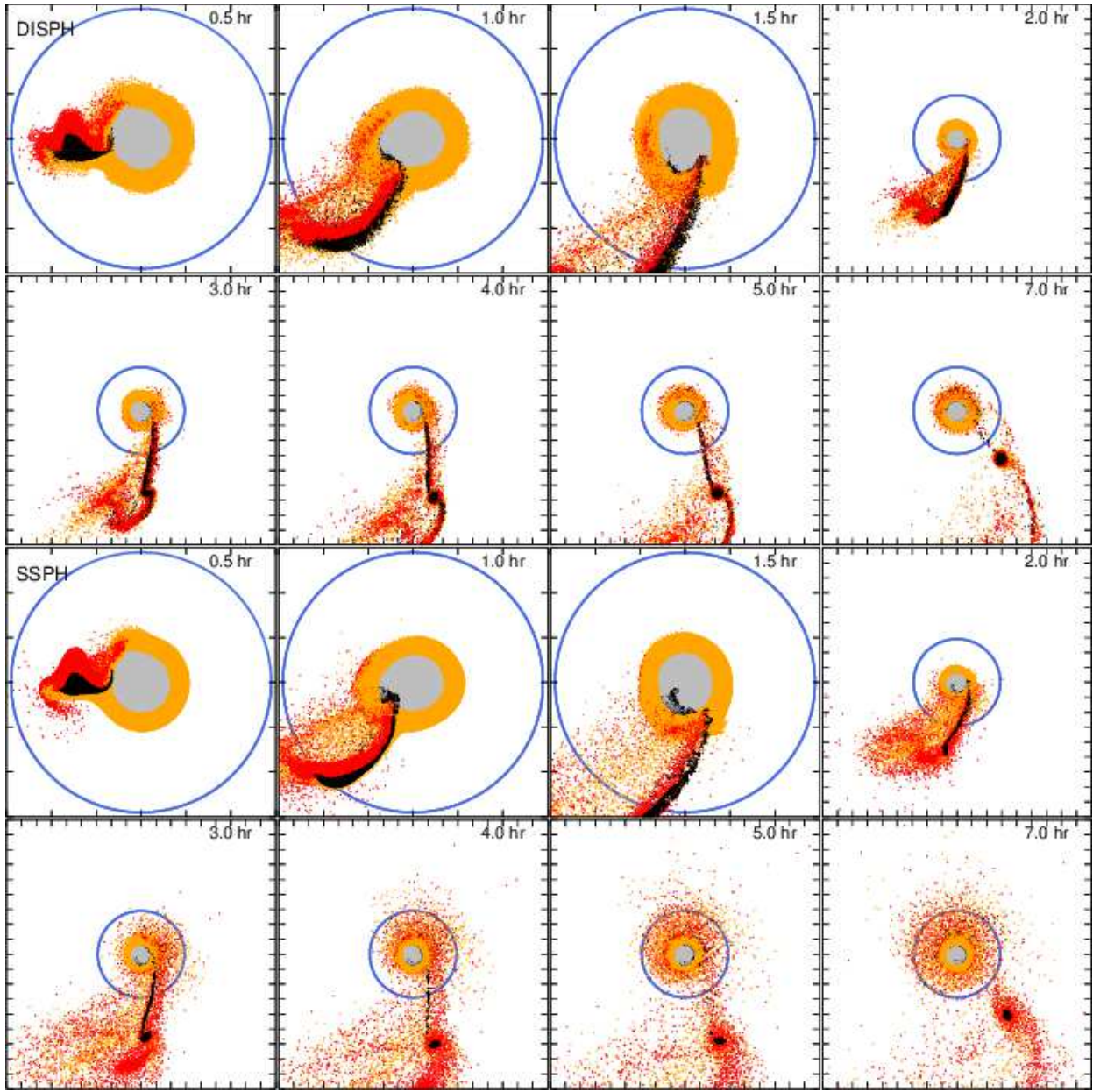}
\caption{
	Same as Fig. \ref {fig:RUN1_X}, but for model 1.32.
}
\label{fig:RUN3_X}
\end{figure}

Figures \ref{fig:RUN4_X} and \ref{fig:RUN3_X} show the same plots as Fig. \ref{fig:RUN1_X} for the models with low-oblique collision and high-oblique collision, respectively.
From these figures, it is obvious that more extended debris disks are greatly formed in the runs with SSPH than in those with DISPH.

\begin{figure}
\plotone{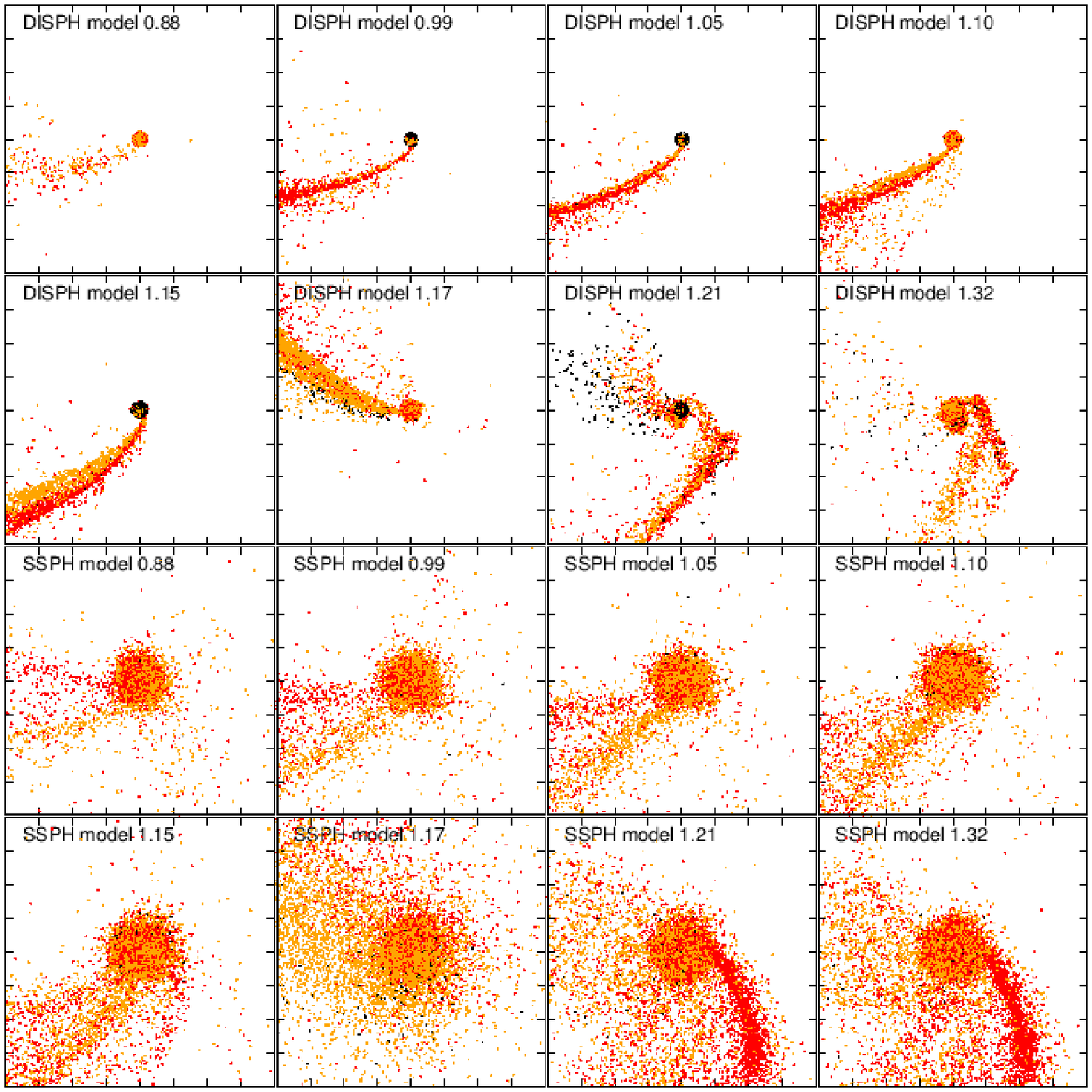}
\caption{
	Snapshots at $t \sim 24$ hrs for all models.
	The panels in upper two rows show the results of DISPH, while those in the lower two rows show those of SSPH.
	The length of each side is $40 R_\mathrm{E}$.
}
\label{fig:diskSnapShots}
\end{figure}

Figure \ref{fig:diskSnapShots} shows the snapshots at the end time of simulations for all runs.
The azimuthal distributions of the disks are different between SSPH and DISPH.
With SSPH, particles are widely distributed.
In particular, in model 1.17, the distribution is nearly azimuthally symmetric.
On the other hand, with DISPH, the distribution of particles is clearly asymmetric and no large disk is formed.

The results of our SSPH runs are similar to those in the previous works.
In both cases, disks extended to outside of the Roche radius are formed. 
On the other hand, the disk is very thin in our DISPH runs for these models.
We quantify this difference more clearly in the section \ref{Sec:SurfaceTension} in order to understand the cause of the difference.

\subsection{Dependence on disk properties on impact angular momentum}\label{Sec:Predicted moon mass}
In this section, we summarize dependence of disk properties on initial impact angular momentum.
The conditions of the successful moon forming impact is defined by 1) the predicted moon mass is comparable to or larger than the present Moon mass, and 2) most of materials of the formed moon comes from the target planet (the proto-Earth).

Figure \ref{fig:Moon_mass}a shows the time evolution of predicted moon mass $M_\mathrm{M}$ for models $0.88, 1.1, 1.15, 1.21$ and $1.32$ with both methods.
Figures \ref{fig:Moon_mass}b and c are the disk specific angular momentum $l_\mathrm{disk}$ and mass $M_\mathrm{disk}$ used to evaluate the moon mass (Eq. \ref{eq:predMoonmass}), which are calculated by setting the coordinate origin at the barycenter of the target's core particle.
Figure \ref{fig:Moon_mass}d shows the time evolution of $M_\mathrm{escape}$ (the escape mass by the impact).
The oscillation in these quantities is due to the post-impact oscillational deformation of the merged planet.
For $L_\mathrm{init} \la 1.15 L_\mathrm{EM}$, $M_\mathrm{M}$ obtained with DISPH is significantly smaller than that with SSPH, because $l_\mathrm{disk}$ is smaller.
For model 1.21 and model 1.32, while $M_\mathrm{disk}$ with DISPH is similar to or rather smaller than that by SSPH, $l_\mathrm{disk}$ is higher for DISPH, resulting in larger $M_\mathrm{M}$ with DISPH than that with SSPH.
In addition, SSPH produces larger amounts of escaping mass than DISPH does.
These result in the larger $M_\mathrm{M}$ with DISPH than with SSPH.

Figure \ref{fig:AMvsMM}a shows the dependance of $M_\mathrm{M}$ on $L_\mathrm{init}$ for runs with SSPH and DISPH.
Generally, $M_\mathrm{M}$ increases with $L_\mathrm{init}$ for both methods as in Fig. \ref{fig:single_comp_AMvsMM} (runs with single-component objects).
Notice that the dependence is much more sensitive in the differentiated objects impacts (Fig. \ref{fig:AMvsMM}a) than in single component objects impacts (Fig. \ref{fig:single_comp_AMvsMM}).
For a high-oblique collision, the impact momentum is transferred to ejecta from the outer parts of the impactor and the target.
The volume of ejecta may be primarily regulated by a geometrical effect if the collision velocity is fixed.
If the volume of ejecta and momentum transferred to the ejecta are the same for a fixed $L_\mathrm{init}$ between an impact of differentiated objects and that of single component objects, the total ejecta mass from differentiated objects is smaller and its post-impact velocity is higher than that from single component objects.
It results in formation of a more spread disk or a hit-and-run collision for the differentiated objects impact at high-oblique impacts.

With DISPH, $M_\mathrm{M}$ is an order of magnitude smaller than those with SSPH for $L_\mathrm{init} \la 1.1 L_\mathrm{EM}$, the trend of which is also found in Fig. \ref{fig:single_comp_AMvsMM}.
On the other hand, in the case of $L_\mathrm{init} \ga 1.15 L_\mathrm{EM}$, DISPH produces larger $M_\mathrm{M}$ than SSPH.
This is because in the runs with SSPH, more materials are ejected during the first contact event than in the DISPH runs (Fig. \ref{fig:Moon_mass}d), probably by the artificial tension at core-mantle boundaries of the impactor and the target, as we discuss in details in section \ref{Sec:SurfaceTension}.
In the panels of relatively high $L_\mathrm{init}$ (models 1.17, 1.21 and 1.32) in Fig. \ref{fig:diskSnapShots} clearly show that much more materials are scattered away in the runs with SSPH, while more compact clumps remain in the runs with DISPH.
As a result, DISPH produces abrupt transition of the predicted moon mass $M_\mathrm{M}$ around $L_\mathrm{init} \sim 1.15 L_\mathrm{EM}$.
Since $M_\mathrm{M}$ is sensitive to the distribution of the disk particles, the difference in $M_\mathrm{M}$ between SSPH and DISPH is more pronounced than that in the distributions of formed disks.

In model 1.21 and model 1.32, the predicted moon mass with DISPH are greater than disk mass (see, Fig. \ref{fig:Moon_mass}a and c).
Since Eq. (\ref{eq:predMoonmass}) is an empirical equation, it is not proper enough for nearly grazing impacts.
The values of $M_\mathrm{M}$ should be treated as a reference value.

Figure \ref{fig:AMvsMM}b shows the material fraction from target in the disk for each model.
These results show that the material fraction from the target is significantly smaller than $90 \%$ in all models with both SSPH and DISPH.
However, in the DISPH runs, $M_\mathrm{M}$ is twice as much as $M_\mathrm{L}$ at $L_\mathrm{init} \ga 1.15 L_\mathrm{EM}$ for $M_\mathrm{imp} = 0.109 M_\mathrm{E}$.
An impact with smaller $M_\mathrm{imp}$ that produces $M_\mathrm{M} \sim M_\mathrm{L}$ should have smaller $L_\mathrm{init}$, while the disk material fraction from the target would be significantly increased.
Because $L_\mathrm{init}$ in that case may be smaller than $L_\mathrm{EM}$, we can adopt higher impact velocity, which may further increase the disk fraction from the target.
We will explore the parameters of the impact with $M_\mathrm{M} \sim M_\mathrm{L}$ and $L_\mathrm{init} \sim L_\mathrm{EM}$ that produces a disk mostly from the target (the proto-Earth) in a separate paper.

\begin{figure}[pt]
\includegraphics[scale=0.85]{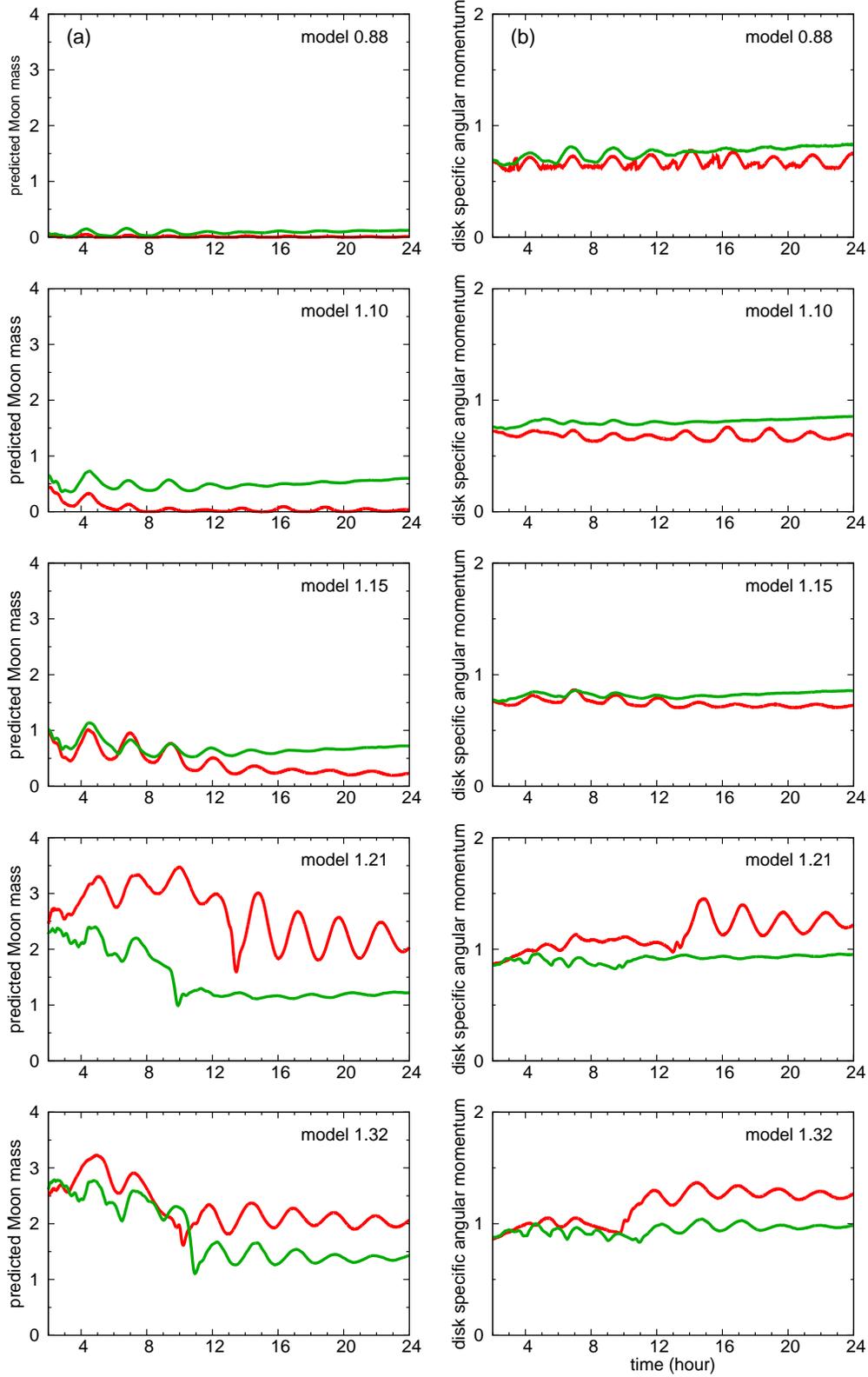}
\caption{
	Time evolution of the predicted moon mass(a), the disk specific angular momentum(b), the mass of disk(c) and escaping mass(d), respectively.
	The red curve indicates the results of DISPH, while green curve indicates those of SSPH.
	The mass and angular momentum are normalized by the current Moon mass, $M_\mathrm{L}$ and $\sqrt{G M_\mathrm{E} R_\mathrm{Roche}}$, respectively.
}
\label{fig:Moon_mass}
\end{figure}

\begin{figure}[pt]
\figurenum{\ref{fig:Moon_mass}}
\includegraphics[scale=0.85]{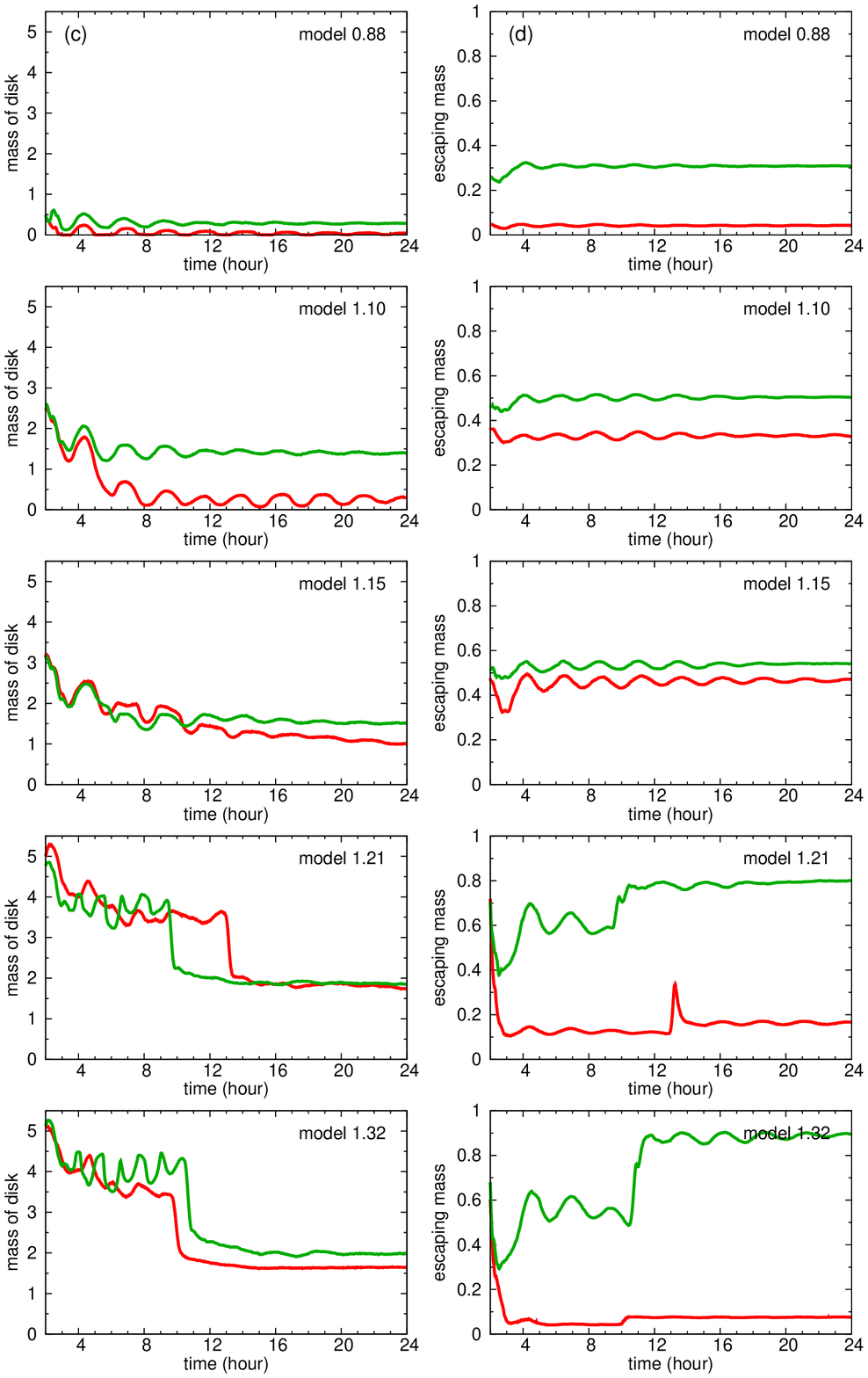}
\caption{
	Continued.
}
\end{figure}

\begin{figure}
\plotone{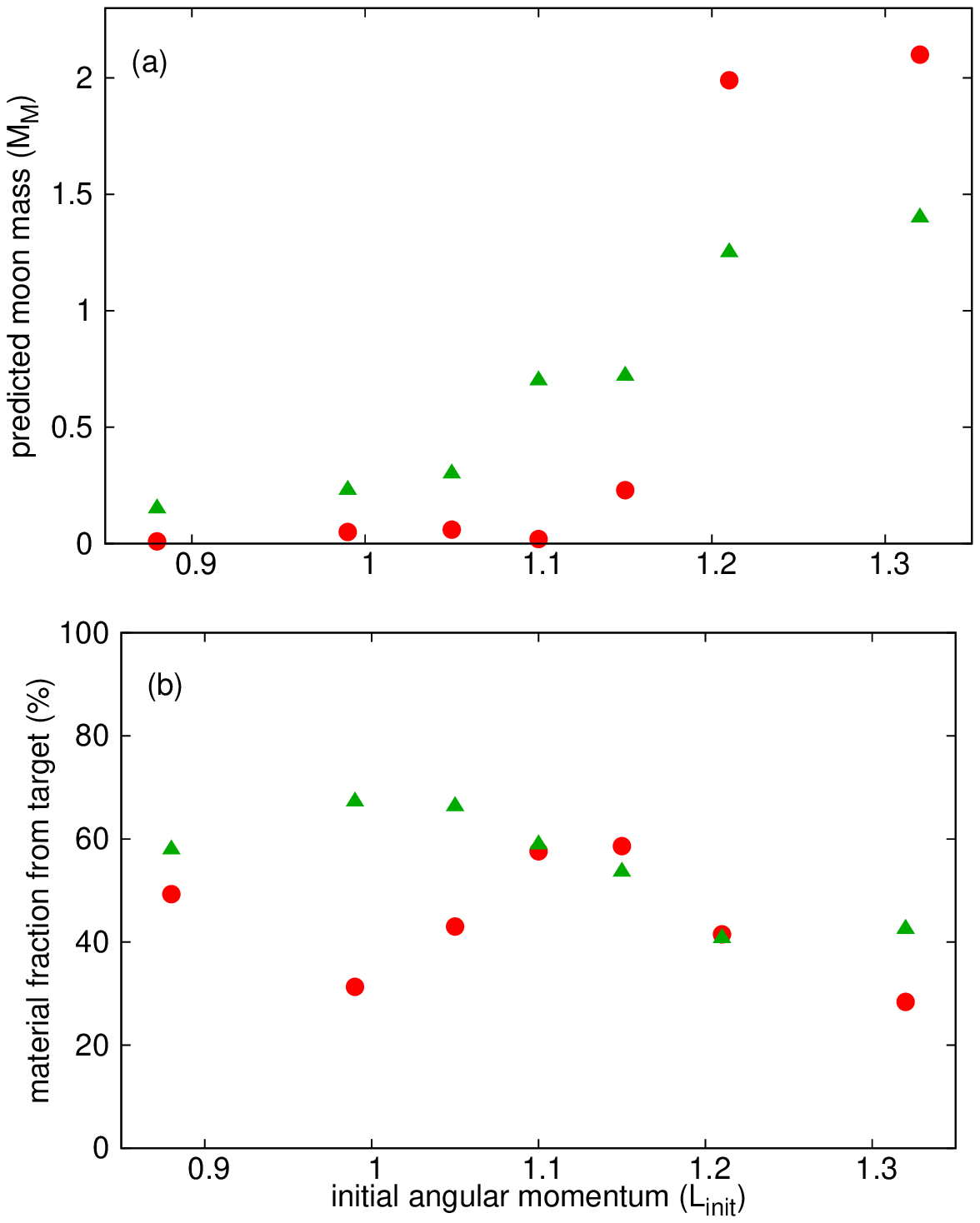}
\caption{
	(a) The initial angular momentum vs. the predicted moon mass with both methods.
	Red circles indicate the results with DISPH, while green triangles indicate those of SSPH.
	The angular momentum is normalized in the current angular momentum of Earth-Moon system.
	(b) The same as (a), but shows target material fraction in disk.
}
\label{fig:AMvsMM}
\end{figure}

\subsection{Effect of the ``Unphysical Surface Tension" in SSPH}\label{Sec:SurfaceTension}
In this section we discuss the origin of the difference between two methods by focusing on the treatment of the core-mantle boundary.

\begin{figure}
\plotone{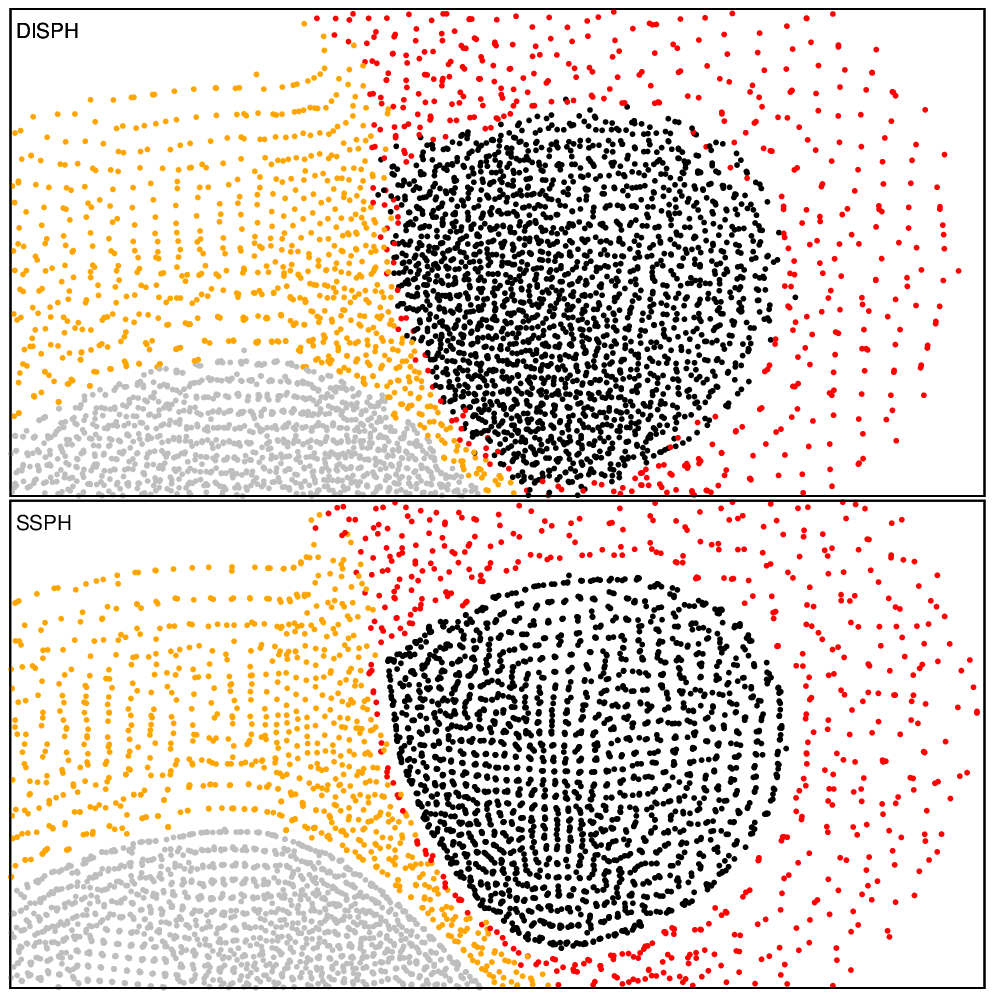}
\caption{
	Close-up views of the impactor of model 1.17 at $t = 6$ min.
	The top panel shows that with DISPH, whereas the bottom panel displays that with SSPH.
	Colors are the same as those of Fig. \ref{fig:RUN1_X}.
}
\label{fig:close_up}
\end{figure}

Figure \ref{fig:close_up} shows the close-up view of SPH particles in model 1.17 at $t = 6$ min.
Clear gaps in the particle distributions are found near the core-mantle boundaries of both the target and the impactor in the SSPH results.
In the case of the DISPH run, such a gap does not exist, and the layers of particles are less clear.
The gap visible in the SSPH run is due to the unphysical surface tension.

\begin{figure}
\plotone{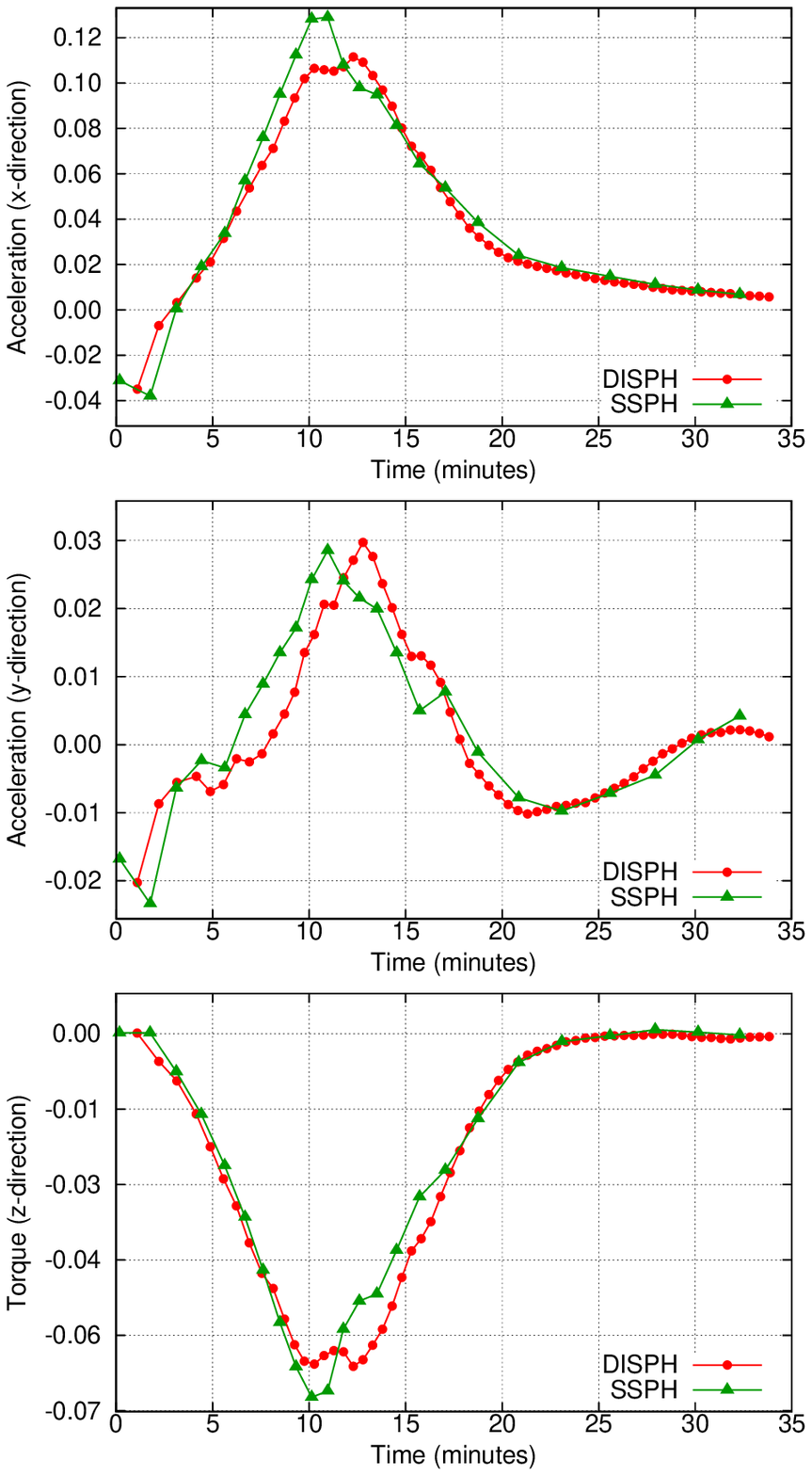}
\caption{
	Time evolution of the mean acceleration along the $x$-direction, $y$-direction and mean torque around the $z$-axis of the impactor's core particle of model 1.17.
	We set $t = 0$ at the time of the first contact of two bodies.
	The green line indicates the value of SSPH while the red line indicates that of DISPH.
	The acceleration and torque are normalized by $G M_\mathrm{E} / R^2_\mathrm{E}$ and $G M_\mathrm{E} / R_\mathrm{E}$, respectively.
}
\label{fig:a_mean}
\end{figure}

In Fig. \ref{fig:a_mean}, we show the acceleration per particle along the $x$-direction, $y$-direction and torque around the $z$-axis of the impactor's core and mantle particles.
The hydrodynamical forces of SSPH and DISPH runs during the impact phase are different.
In the first 10 minutes, the accelerations in both directions of SSPH are larger than those of DISPH.
This difference results in the gap shown in Fig. \ref{fig:close_up}.
From $t = 12$ - $17$ minutes, the SSPH result shows larger torque than that of DISPH.

\begin{figure}
\plotone{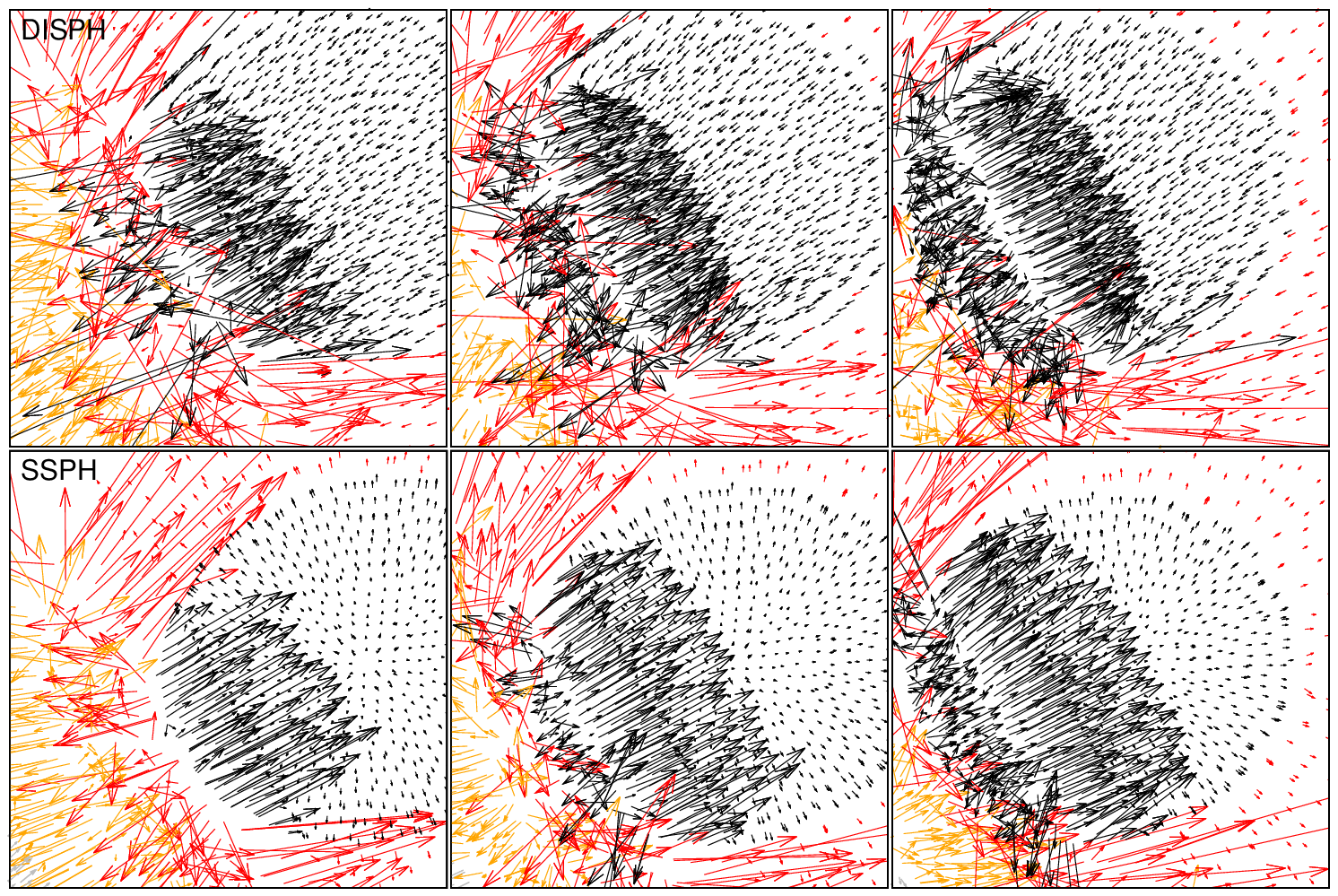}
\caption{
	The hydrodynamical forces externes on individual particles of the impactor.
	The upper and lower panels show the DISPH and SSPH runs, respectively.
	The snapshots are about $3, 4$ and $5$ min after the initial contact of the two bodies from left to right.
	Colors of particles are the same as those in Fig. \ref{fig:RUN1_X}.
	The length of the arrows are proportional to the absolute values of hydrodynamical forces.
}
\label{fig:accel}
\end{figure}

Figure \ref{fig:accel} illustrates the effect of this surface tension.
In the lower left panel (SSPH, $t = 3$ min), none of the impactor's core particle suffers negative $y$-directional force, while in the corresponding snapshot with DISPH at $t = 3$ min (the upper left panel), some particles suffer negative $y$-directional force.
The amplitude of the acceleration of impactor's core particles is much larger for the SSPH run.
Thus, the impactor particles gain upward velocity (in the direction of $y$-axis), while losing the forward velocity (negative direction of $x$-axis), compared to the DISPH run.
It is most likely that this difference is due to the numerical error of SSPH at the contact density discontinuity (the core-mantle boundary) and it results in the difference in the formed disks between the DISPH and SSPH runs.

\section{Summary}\label{Sec:Summary}
The giant impact (GI) is the most accepted model for the origin of the Moon.
However, it is now being challenged.
The identical isotope ratios between the Earth and the Moon found by recent measurement require a survey of new ranges of impact parameters, because the impact previously referred to as a ``successful Moon forming impacts" produces a moon mostly consisting of materials from the impactor rather than those from the proto-Earth.

We have re-investigated GI by newly developed ``Density Independent SPH'' (DISPH) scheme with Tillotson EOS.
Recently it is recognized that the standard SPH scheme (SSPH) has a serious problem in the treatment of contact discontinuities because SSPH assumes differentiability of density.
The core-mantle boundary is a contact discontinuity and the planetary surface (free surface) also has a density jump.
The errors result in the unphysical surface tension around the contact discontinuity and the free surface.
Since DISPH assumes differentiability of pressure instead of density, it can properly treat the core-mantle boundary, although the treatment of the free surface is not significantly improved from SSPH, compared with the contact discontinuity.
Several tests of DISPH in Appendix A, B, C and D show advantage to and compatibility with SSPH.

Recent studies have pointed out that the standard SPH scheme (SSPH) has a serious problem in the treatment of contact discontinuities.
The errors result in the unphysical surface tension around the contact discontinuity and the free surface.

\citet{H+13} extended \citet{SM13}'s Density Independent SPH (DISPH) scheme to non-ideal EOS such as Tillotson EOS, as summarized in section 2.1.
We have compared the results between DISPH and SSPH, focusing on properties of circumplanetary disks generated by GI.
To distinguish between the effects of the core-mantle boundary and the free surface, we performed simulations of the collisions between two single component objects and those between differentiated objects with core-mantle structure.

In the case of collisions between single-component objects, compared with SSPH, DISPH always produces more compact disks, for which smaller moon masses are predicted.
This is because numerical repulsive force appears around the free surface, in SSPH runs (section \ref{Sec:single comp}).
Note that since the predicted moon mass is sensitively dependent on the distribution of the disk particles, slight difference in the distribution between SSPH and DISPH can result in significant difference in the predicted moon mass.

On the other hand, in the case of collisions between differentiated objects with core-mantle boundary, DISPH predicts more massive moon masses than SSPH does for high-oblique impacts, while it still predicts lower mass moons for low-oblique impacts (section \ref{Sec:time evolve}).
The different dependence on the initial impact angular momentum from the single component objects would come from the transfer of impact momentum to the mantle layer with low density and numerical repulsive force at the core-mantle boundary (section \ref{Sec:time evolve} and \ref{Sec:Predicted moon mass}).
The overall trend that the predicted moon mass increases with the initial impact angular momentum is common between SSPH and DISPH.

Note that our result is consistent with the conclusion by \citet{C+13}: SSPH and a grid code, AMR, produce disks that predict similar moon mass.
They did a comparison for impacts with parameters similar to model 1.21.
As we showed in Fig. \ref{fig:AMvsMM}, for model 1.21, SSPH and DISPH show similar results within 50\% of the predicted moon mass.
Thereby, DISPH produces consistent results with AMR for this parameter.
The comparison with grid codes is necessary for other initial impact angular momentum for which DISPH and SSPH significantly differ from each other.
However, clump structure looks different among AMR, SSPH and DISPH, suggesting that the angular momentum distributions are different among them.

What we want to stress in this paper is that properties of circumplanetary disks generated by GI are sensitive to the choice of the numerical scheme.
Only the difference in the treatment of a contact discontinuity (core-mantle boundary) between DISPH and SSPH significantly affects the results.
Other effects such as the treatment of free surface, shock propagation, heating so on are also likely to change the results.
The results of GI also depend on the EOS, the initial thermal structure, density profiles, material strength and numerical resolution and so on.

We need be very careful when some conclusions are drawn from the numerical simulations for GI, because planets consist of solid layers with different compositions but not uniform gas and current numerical schemes have not been developed enough to treat planets.
Thus, we need to develop numerical codes suitable for GI between planets, step by step.
The next step of DISPH would be handling of free surfaces and shock propagation that currently has a free parameter $\alpha$.
For GI, code-code comparisons are now needed.
Comparison to experiments or other numerical schemes in the case of simple impact problem is also needed to calibrate the code.
These are left for future works.

\acknowledgments
\section*{acknowledgement}
The authors thank Matthieu Laneuville, Prof. Melosh and the anonymous referees for giving us helpful comments on the manuscript.
This work is supported by a grant for the Global COE Program, `From the Earth to ``Earths''', MEXT, Japan.
Part of the research covered in this paper research was funded by MEXT program for the Development and Improvement for the Next Generation Ultra High-Speed Computer System, under its Subsidies for Operating the Specific Advanced Large Research Facilities.
It was also supported in part by a Grant-in-Aid for Scientific Research (21244020), the Grant-in-Aid for Young Scientists A (26707007) and Strategic Programs for Innovative Research of the Ministry of Education, Culture, Sports, Science and Technology (SPIRE).

\appendix
\section{3D shock tube problem by DISPH with $\alpha$}\label{Sec:App1}
To study the effect of choice of a parameter $\alpha$ in the DISPH scheme, we performed two calculation 3D shock tube problems with SSPH and DISPH with varying the parameter $\alpha$.
One calculation is with the ideal gas EOS and another is with Tillotson EOS.
The parameter $\alpha$ is taken to be $1.0$, $0.5$ and $0.1$.

The initial condition of the 3D shock tube problem for ideal gas EOS is set as follows:
\begin{eqnarray}
(\rho, p, u) = \left\{ \begin{array}{ll}
	(1.0, 1.0, 2.5) & (x < 0.5), \\
	(0.5, 0.2, 1.0) & (\mathrm{otherwise}).
\end{array} \right.
\end{eqnarray}
The velocity of each side is set to be $0$.
We employ a 3D computational domain, $0 \leq x < 1$, $0 \leq y < 1/8$ and $0 \leq z < 1/8$, and periodic boundary conditions are imposed in all directions.
We employ ideal gas EOS with specific heat ratio $\gamma = 1.4$.
The particle separation of high density side is set to $1/512$; thus, the total number of particles is $1572864$.

The initial condition of the 3D shock tube problem for Tillotson EOS is set as follows:
\begin{eqnarray}
(\rho, p, u) = \left\{ \begin{array}{ll}
	(2.0, 7.0, 2.33) & (x < 0.5), \\
	(1.0, 3.5, 4.85) & (\mathrm{otherwise}).
\end{array} \right.
\end{eqnarray}
The parameters of granite are adopted for Tillotson EOS.
The density and specific internal energy are normalized in reference density $\rho_0$ and reference energy $E_0$ \citep[][]{T62, M89}.
The computational domain and number of particles are the same as those in the calculation with the ideal gas EOS.

\begin{figure}
\plotone{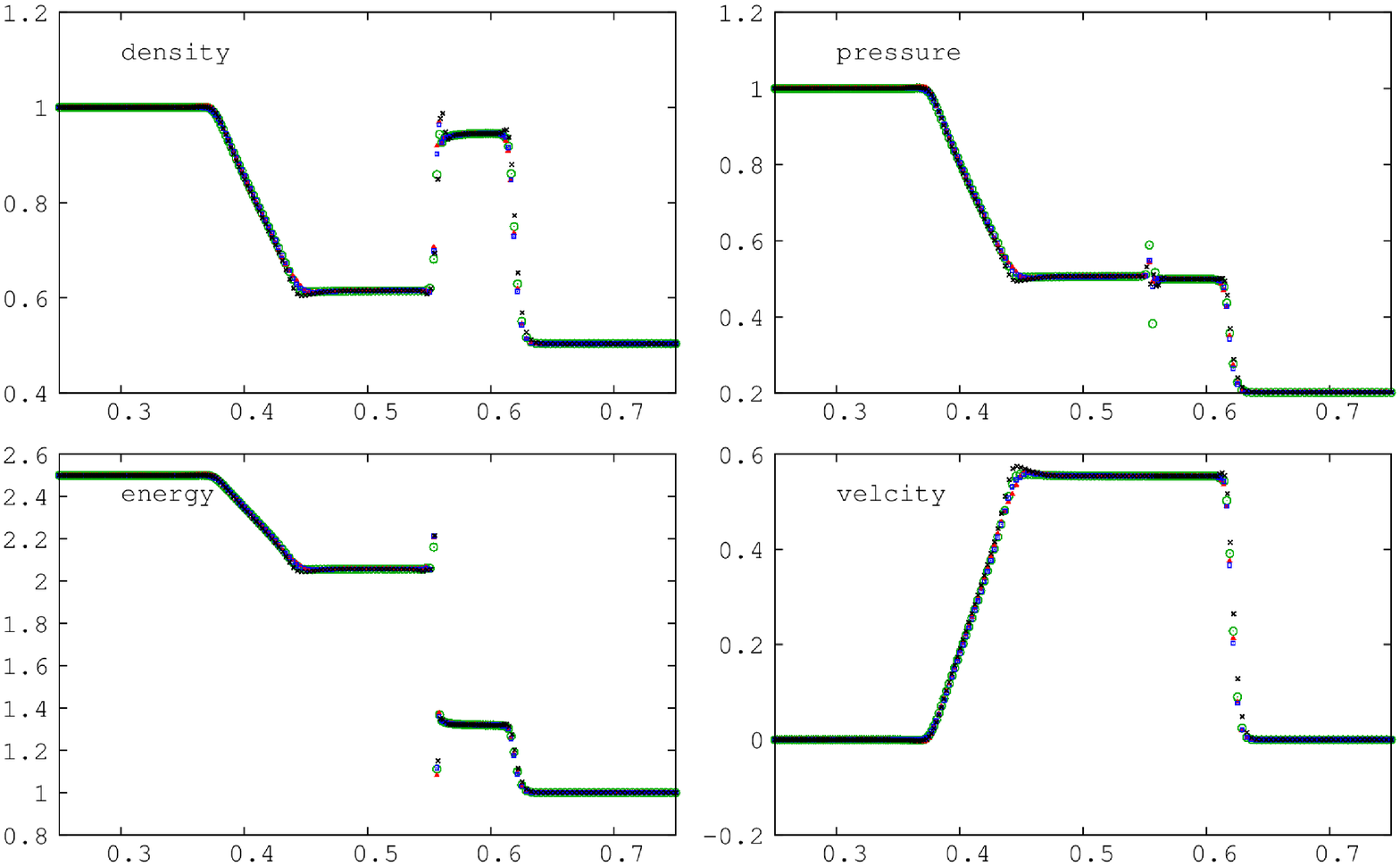}
\caption{
	Snapshot from the 3D shock tube problem with both DISPH and SSPH.
	Density, pressure, velocity and specific internal energy are shown.
	Green circles, red triangles, blue squares and black crosses indicate the results with SSPH, DISPH with $\alpha = 1.0$, DISPH with $\alpha = 0.5$ and DISPH with $\alpha = 0.1$, respectively.
}
\label{fig:st3D}
\end{figure}

\begin{figure}
\plotone{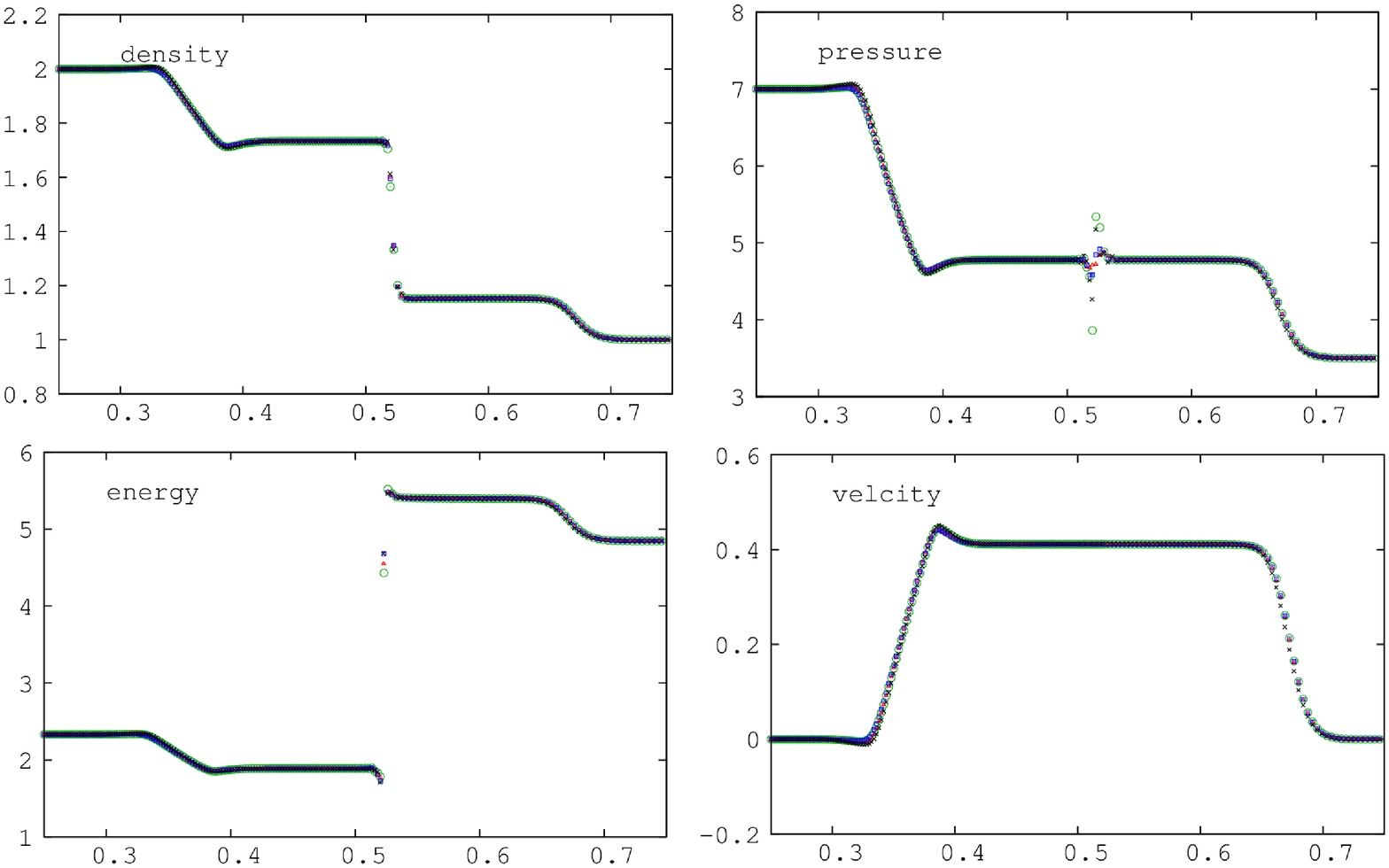}
\caption{
	Same as Fig. \ref{fig:st3D}, but with Tillotson EOS.
}
\label{fig:st3Dt}
\end{figure}

Figure \ref{fig:st3D} shows snapshots at $t = 0.1$ of the 3D shock tube problem for the ideal gas EOS with both SSPH and DISPH.
Both methods show similar results, except for the contact discontinuity (at $x \sim 0.55$).
As expected, DISPH shows fairly smaller pressure blips than SSPH at the contact discontinuity for all values of $\alpha$, while larger jumps are found in density with DISPH.
These results are consistent with those shown in \citet{SM13} and \citet{H+13}.
Figure \ref{fig:st3Dt} shows snapshots at $t = 0.05$ of the 3D shock tube problem for Tillotson EOS with both SSPH and DISPH.
Also in this case, DISPH shows smaller pressure blips around the contact discontinuity.
For all values of $\alpha$, DISPH shows better treatment of the contact discontinuity.
The dependence of the magnitude of the pressure blip on $\alpha$ is different between the ideal gas EOS and Tillotson EOS.
With Tillotson EOS, the pressure blip is higher for smaller value of $\alpha$ while the blip is still smaller than that with SSPH.
However, as shown below, the treatment of a free surface is better for smaller $\alpha$ even with DISPH. 

Note that this pressure blip can be a serious problem when we treat the contact discontinuity.
\citet{SM13} and \citet{H+13} showed the consequence of this pressure blip \citep[see Fig. 4 in][]{H+13}.
In their hydrostatic equilibrium tests, they put the high-density square in the low-density ambient in the pressure equilibrium.
With SSPH, the high density square, which should remain its initial shape, quickly transforms into circular shape.
With DISPH, on the other hand, the high density square remains its initial shape.
This means that with SSPH, simulations suffer from the unphysical momentum transfer.
\citet{SM13} and \citet{H+13} also showed the results of Kelvin-Helmholtz instability test, in which the contact discontinuity plays very important role \citep[see Fig. 5 in][]{H+13}.
As expected, SSPH shows unphysical surface tension effect, while DISPH clearly eliminates it.
Previous simulations should have suffer from this unphysical effect.

The treatment of the region with abrupt change in the pressure that corresponds to a free surface is improved by DISPH, in particular with small value of $\alpha$.
To show this, we show the pressure field around the strong shock region with Tilloston EOS.
The initial pressure distribution is set as follows:
\begin{eqnarray}
p = \left\{ \begin{array}{ll}
	10^{6} & (x < 0.5), \\
	1.0 & (\mathrm{otherwise}).
\end{array} \right.
\end{eqnarray}
The density is uniformly set to be $\rho_0$.
Figure \ref{fig:strong_shock} shows the pressure field and the error at the very first step, where the error is defined as follows:
\begin{eqnarray}
\Delta p_i = \frac{|p_i - p|}{p}.
\end{eqnarray}
This figure clearly shows that taking the parameter $\alpha$ small improves the treatment of the large pressure jump, such as free surface.

In Fig. \ref{fig:AMvsMM_w1}, we show the results of three runs of GI with $\alpha = 1$.
The results are somewhat different from those with $\alpha = 0.1$.
However, just like the case with $\alpha = 0.1$, DISPH produces a smaller moon mass in the model 1.10 and a larger moon mass in the model 1.32.

Our modification for DISPH has good capability for both the shock and contact discontinuity, although DISPH includes the free parameter $\alpha$.
The tests of 3D shock tube and a free surface with ideal gas EOS show that $\alpha = 0.1$ may be the best choice, similar to \citet{SM13}.
However, with Tillotson EOS, different dependence on $\alpha$ can be seen.
The results for GI with Tillotson EOS shows slightly different dependence on α, though the number of runs is few.
Thus, more careful calibration for the dependence on $\alpha$ should be done, which is left for future work.
Unless otherwise specified, in the following, we adopt $\alpha = 0.1$.

\begin{figure}
\plotone{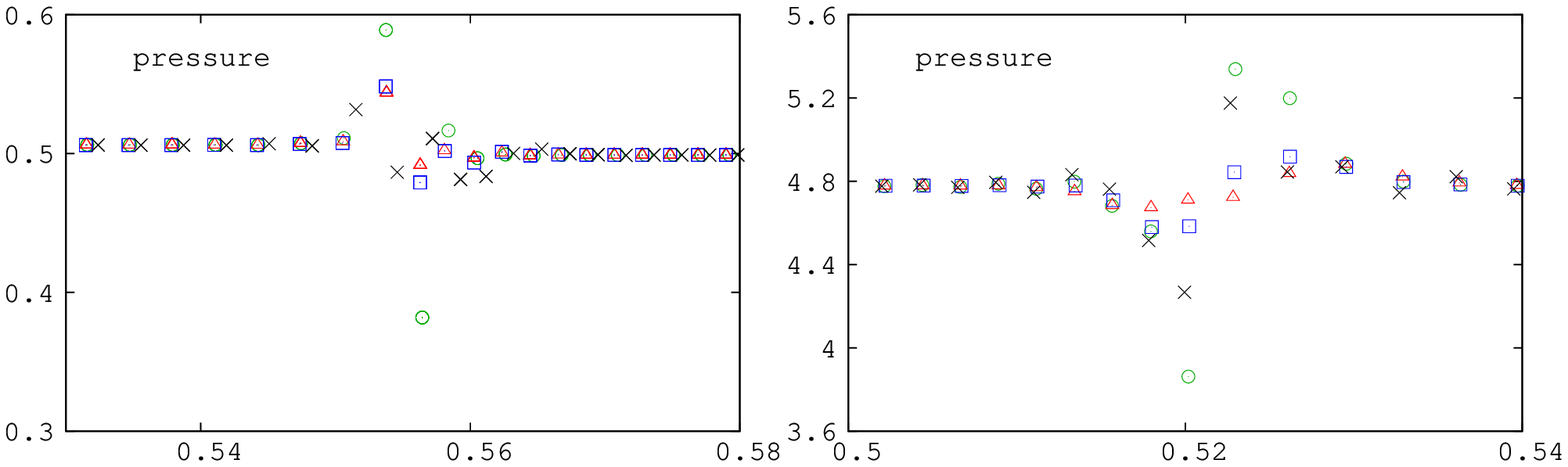}
\caption{
	Close-up around the contact discontinuity of Figs. \ref{fig:st3D} and \ref{fig:st3Dt} from left to right.
}
\label{fig:st_cu}
\end{figure}

\begin{figure}
\plotone{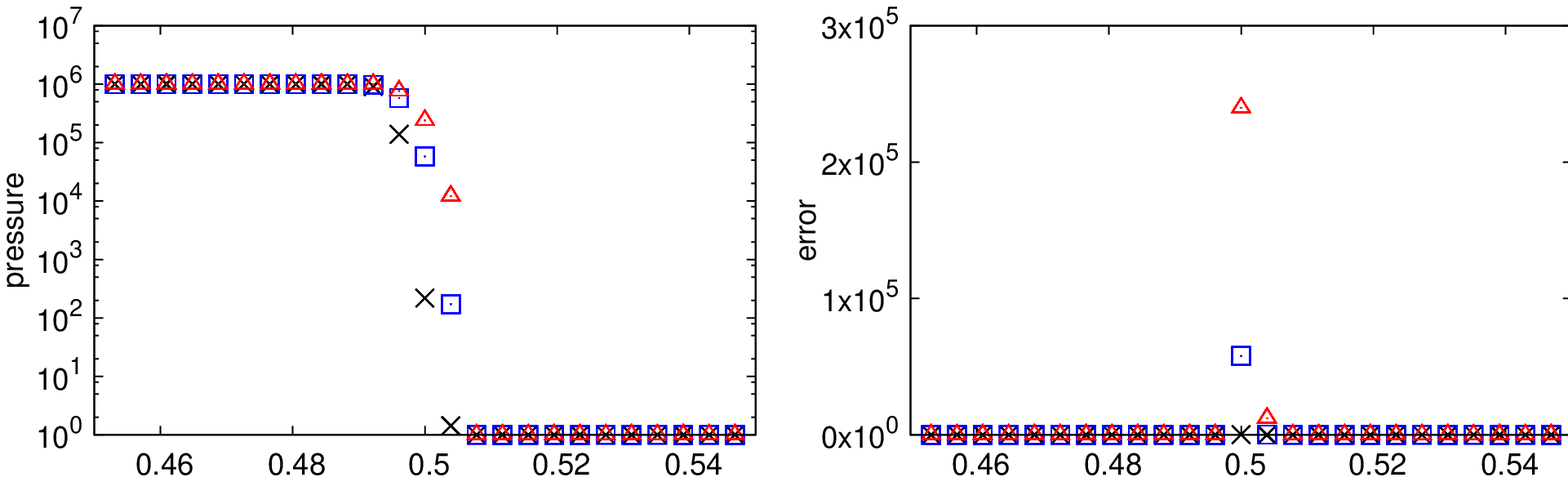}
\caption{
	The left column shows pressure around the strong shock for Tillotson EOS with DISPH with $\alpha = 1.0, 0.5$ and $0.1$, respectively.
	The right column shows the relative error of the pressure.
	The colors of symbols are the same as Fig. \ref{fig:st3D}.
}
\label{fig:strong_shock}
\end{figure}

\begin{figure}
\plotone{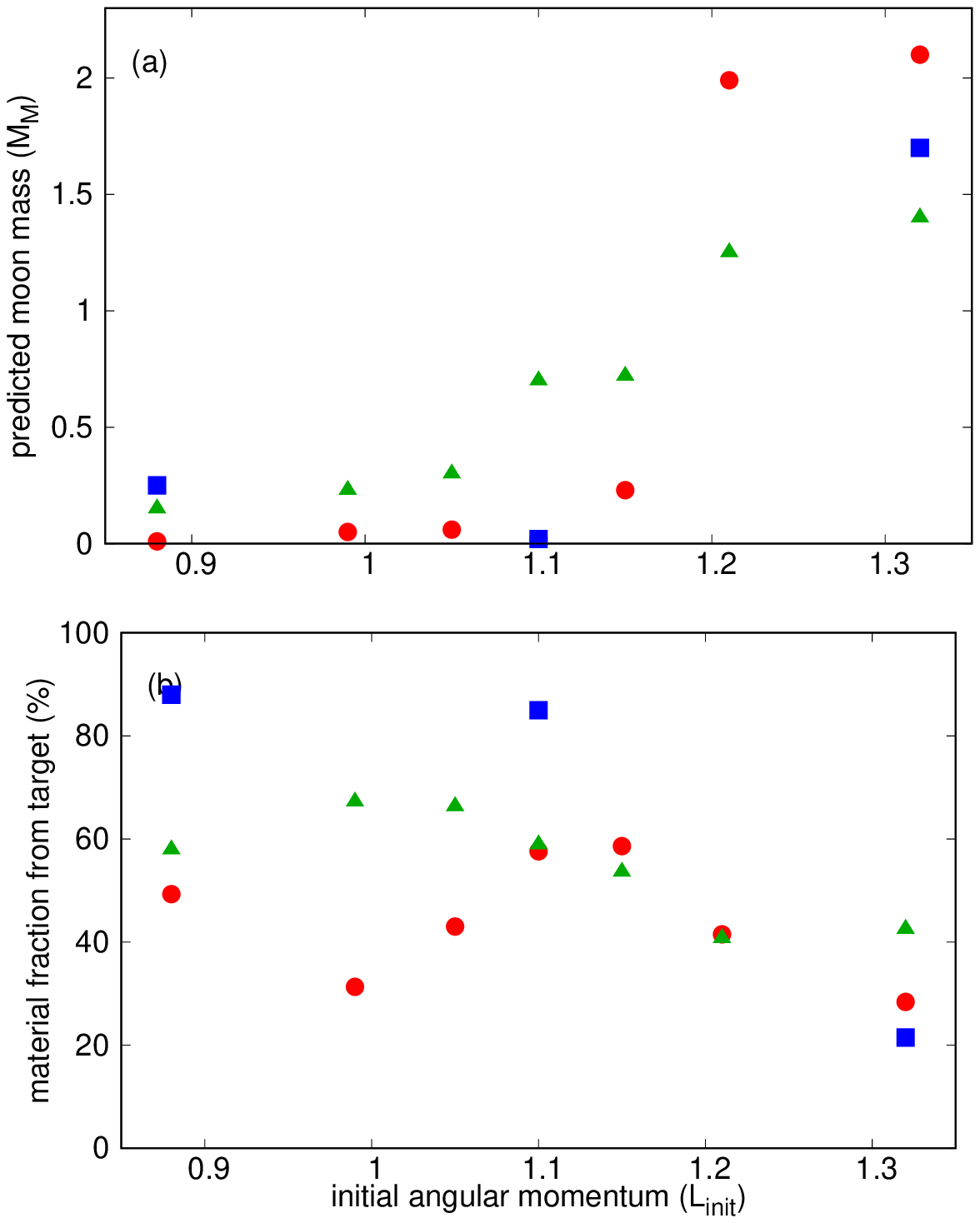}
\caption{
	The same as Fig. \ref{fig:AMvsMM}, but shows the results with $\alpha = 1.0$ (blue squares).
}
\label{fig:AMvsMM_w1}
\end{figure}

\section{Tests for the conservation of angular momentum}\label{Sec:App2}
For the calculation of GI problems, it is important to treat the angular momentum transfer correctly.
In both SSPH and DISPH, since the interaction between two particles is pairwise, the global angular momentum is conserved.
However, it is often said that SSPH does not treat the local angular momentum transfer correctly; unphysical angular momentum transfer due to the so-called zeroth order error and spurious viscosity appear.
In this appendix, to test whether SSPH and DISPH can treat the angular momentum transfer correctly or not, we performed two well-posed tests; one is the Keplerian disk test \citep[e.g.,][]{H15, CD10} and the other is the Gresho vortex test \citep[e.g.,][]{GC90}.

For the Keplerian disk test, we initialize two-dimensional disk whose surface density is set to $1.0$.
The inner and outer edges of the disk are set to $0.5$ and $2.0$.
The initial pressure of the disk is set to $10^{-4}$ and the heat capacity ratio of the ideal gas is set to $1.4$.
The self gravity between particles are ignored, while the gravity from the central star $\vec{a} = - \vec{r} / r^3$ acts on each particle.
In this test we employ $48228$ particles in total.

For the Gresho vortex test, we employ $[-1, 1)^2$ periodic boundary computational domain with the uniform density of unity.
The initial pressure and azimuthal velocity distributions are as follows:
\begin{eqnarray}
p(R) & = & \left\{ \begin{array}{ll}
	5 + 12.5 R^2 & (R < 0.2), \\
	9 + 12.5 R^2 - 20 R + 4 \log(5 R) & (R < 0.4), \\
	3 + 4 \log(2) & (\mathrm{otherwise}),
\end{array} \right.\\
v_\phi(R) & = & \left\{ \begin{array}{ll}
	5R & (R < 0.2), \\
	2 - 5R & (R < 0.4), \\
	0 & (\mathrm{otherwise}),
\end{array} \right. \label{eq:velocity_profile}
\end{eqnarray}
where $R = \sqrt{x^2 + y^2}$.
In this test we employ $16,384$ particles in total.

\begin{figure}
\plotone{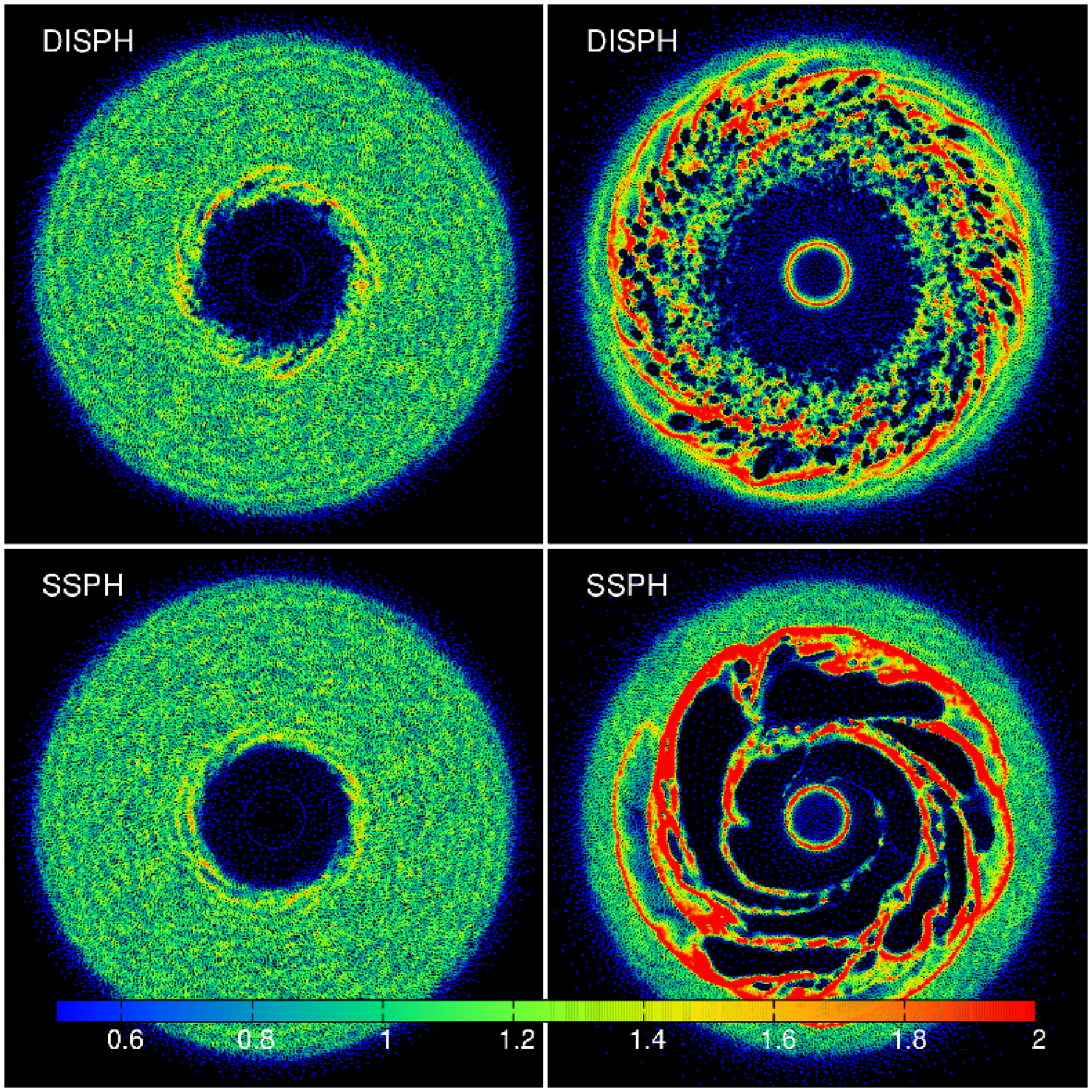}
\caption{
	Snapshots of the Keplerian disk test at $t = 3.5$ and $10$.
	The times are normalized by the orbital time at $r = 1$.
	The top two panels show the results with DISPH, while bottom two panels show those of SSPH.
	The color contour is the density.
}
\label{fig:disk}
\end{figure}

\begin{figure}
\plotone{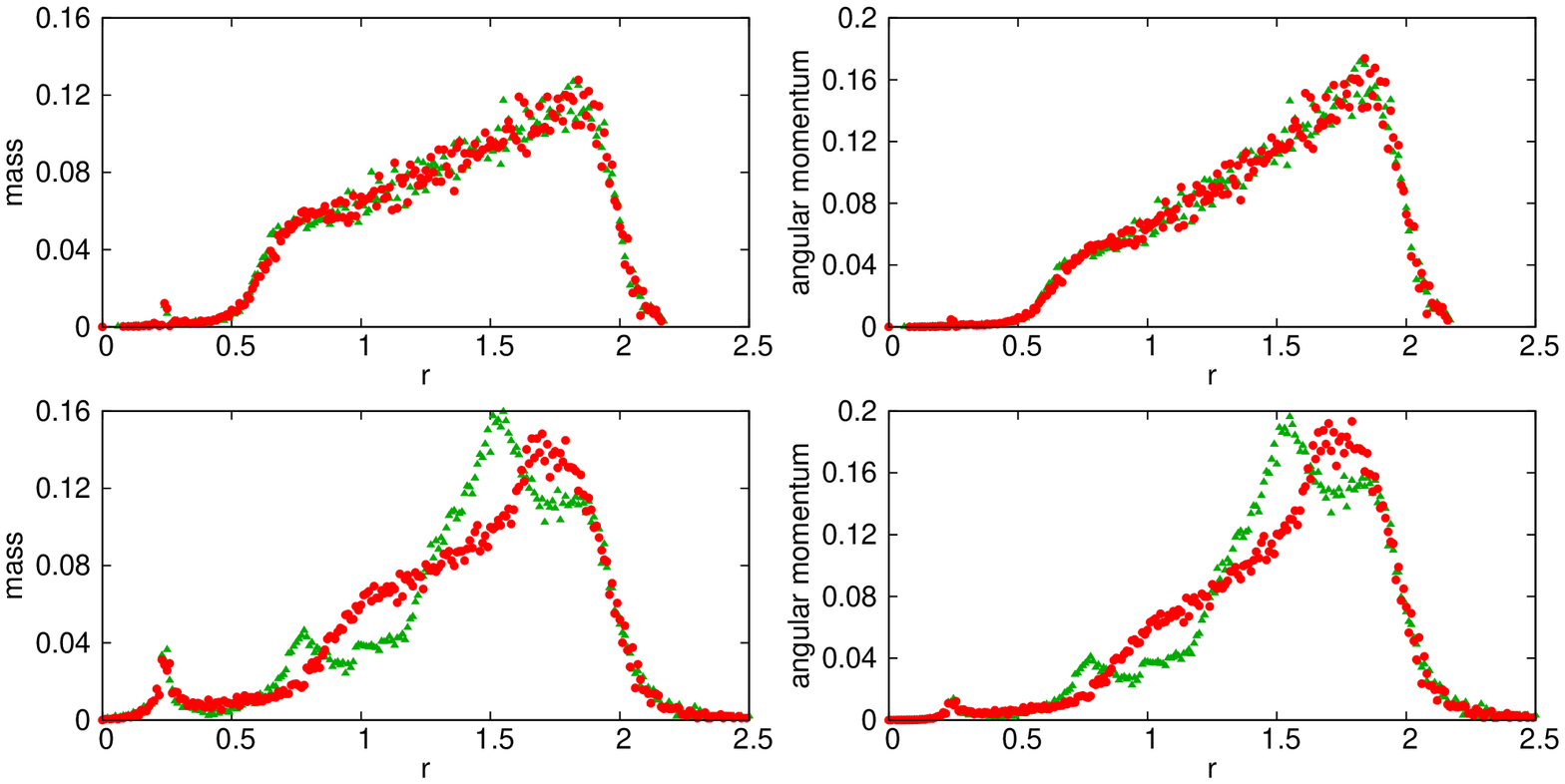}
\caption{
	Binned mass and angular momentum of the Keplerian disk test.
	The top two panels shows the results at $t = 3.5$, while bottom two panels shows those at $t = 10$.
	The times are normalized by the orbital time at $r = 1$.
	Red circles indicate the result with DISPH, while green triangles indicate that of SSPH.
}
\label{fig:bin}
\end{figure}

Figure \ref{fig:disk} shows the snapshots of the Keplerian disk test with both methods.
Neither is accurate enough; the disk breaks up less than $10$ orbits.
At the time $t = 3.5$ orbits, both methods show quite similar results, except the inner edge $r \simeq 0.5$.
Then, SSPH shows catastrophic break up of the rotating disk and makes large filament-like structure, similar to the previous studies \citep[e.g.,][]{H15}.
On the other hand, DISPH also shows break up of the rotation disk.
However, DISPH does not produce filament-like structure, though the break up of the disk can be seen.
With SSPH, the outer regions of the disk is still remain while with DISPH virtually all regions are distorted.
We also note that the radial distributions of the mass and angular momentum are similar between two methods at the time $t = 3.5$ orbits (see, Fig. \ref{fig:bin}).
However, as expected, the radial distributions of the mass and angular momentum are quite different at the time $t = 10$ orbits.

\begin{figure}
\plotone{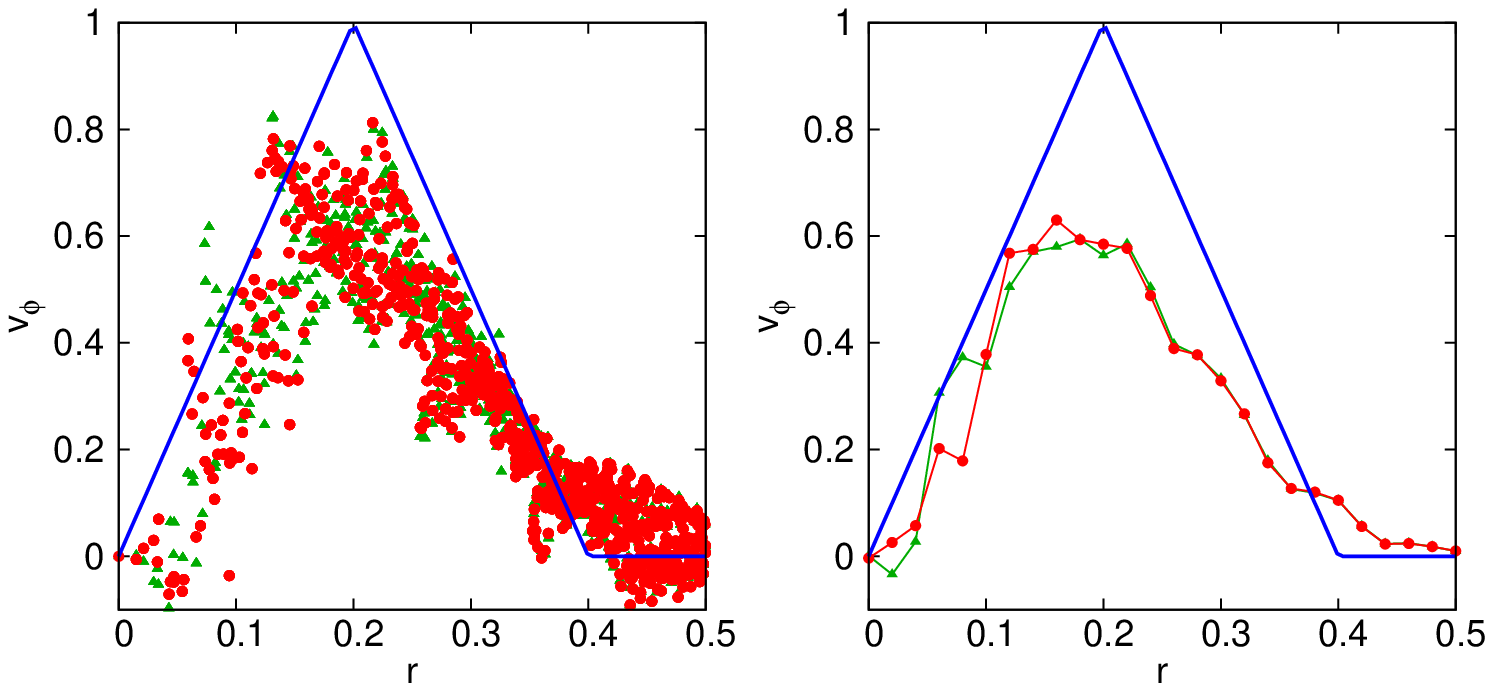}
\caption{
	Radius vs. azimuthal velocity from the Gresho vortex problem at $t = 1.0$.
	Red circles indicate the result with DISPH, while green triangles indicate that of SSPH.
	The solid blue line is the ideal velocity profile (see, Eq. \ref{eq:velocity_profile}).
	The left panel shows the azimuthal velocity of each particle, while the right panel shows binned value with binned width $0.02$.
}
\label{fig:Gresho}
\end{figure}

Figure \ref{fig:Gresho} shows the results of the Gresho vertex test with both methods.
Both methods show similar results; substantial velocity noise appears, similar to the previous studies \citep[e.g.,][]{S10, H15}.

From these tests, we should conclude that both methods can handle the local angular momentum transfer to the same degree.
Note that in GI simulations, we set the end time at $t = 24$ hrs, which corresponds to about $3.4$ orbital time at the Roche limit.
Figure \ref{fig:disk} shows that both methods can treat the local angular momentum transfer until $3.4$ orbital time.
Overall, both SSPH and DISPH are capable of dealing with rotation disks with similar degree, as far as the simulation time is less than $3.4$ orbital periods.

\section{Sedov-Taylor blast wave test}\label{Sec:App4}
Here we show the results of Sedov-Taylor blast wave test, which shows the capability of the strong shock.
The initial condition of this test is the same as that used in \citet{SM13}.
We employ 3D computational domain $[0, 1)^3$.
We first place equal-mass $2,097,152$ particles in the glass distribution with the uniform density of unity.
Then, the explosion energy is added to the central $32$ particles.
In this test we use the ideal gas EOS with $\gamma = 5/3$.

\begin{figure}
\plotone{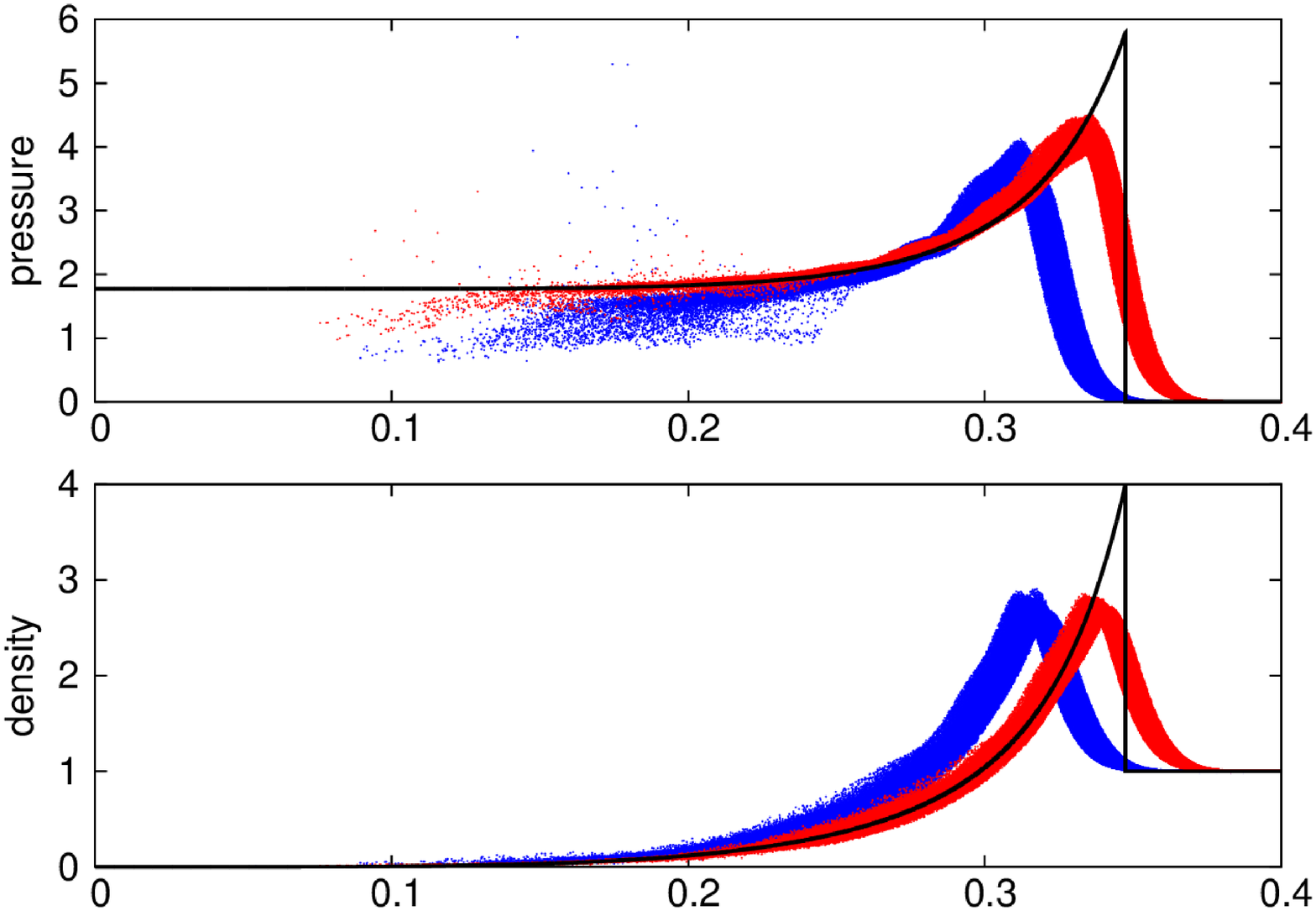}
\caption{
	Radius vs. pressure and density of the Sedov-Taylor blast wave test at $t = 0.05$ are shown in upper and lower panel, respectively.
	The red dots indicate the results with the grad-$h$ term, while blue dots indicate those without the grad-$h$ term.
	The solid line indicates semi analytic solution.
}
\label{fig:sedov}
\end{figure}

Figure \ref{fig:sedov} shows the results of this test with DISPH with and without grad-$h$ term.
Without grad-$h$ term, clearly, the shock propagates slower than the semi analytic solution.
DISPH with the grad-$h$, on the other hand, shows better results than DISPH without the grad-$h$ term.
Our DISPH with the grad-$h$ can treat the strong shock.
Note that these results are consistent to \citet{SM13} and \citet{H13}.

\section{Test for the treatment of the free surface}\label{Sec:App3}
To check the capability for the free surface, we performed three simple 2D tests which include free surface with both DISPH and SSPH.
The first test is hydrostatic equilibrium test, which is carried by \citet{M94}.
The second test is the vertical impact of aluminium-to-aluminium test and the third test is the glass-on-water test.
The latter two tests are performed by \citet{P+08}.

For the first test, we employ 2D computational domain $[0, 40) \times [0, 40)$ and $4,096$ particles in total.
The periodic boundary condition is imposed on the $x$-direction.
We set up the fluid which is initially in pressure equilibrium under the constant gravitational acceleration $g = - 10$ along the $y$-direction.
We fix the positions and internal energies of all particles with $y < 4$.
Around $y = H$, there is a free surface.
In this test we use the following EOS for linear elastic material:
\begin{eqnarray}
p = A \left(\frac{\rho}{\rho_0} - 1 \right).
\end{eqnarray}
The material parameters $A$ and $\rho_0$ are set to the granite's values in the Tillotson EOS.
Since this system is in a hydrostatic equilibrium, particles should maintain their initial positions.

For the second test, we first placed the target particles in $[0, 10^4)^2$.
Then, we placed projectile particles with the impact velocity $20$ km/sec.
We employ $67,597$ particles in total.
We set the impact angle to $0^\circ$ (vertical impact).
The radius of projectile is set to $10^3$ m.
In this test we use the Tillotson EOS and the material parameters are set to the aluminium's values, similar to \citet{P+08}.
Note that in this test, we omit the material strength.

In the third test, we followed the early time evolution of the glass-on-water test, following \citet{P+08}.
We first place target particles in $[0, 0.05)^2$.
Then, we placed projectile particles with the impact velocity $4.64$ km/sec.
We employed $65,336$ particles in total.
We set the impact angle to $0^\circ$ (vertical impact)
The radius of projectile is set to $1$ mm.
We used the Tillotson EOS and the material parameters are set to water for target and wet tuff for projectile \citep{M89}.

\begin{figure}
\plotone{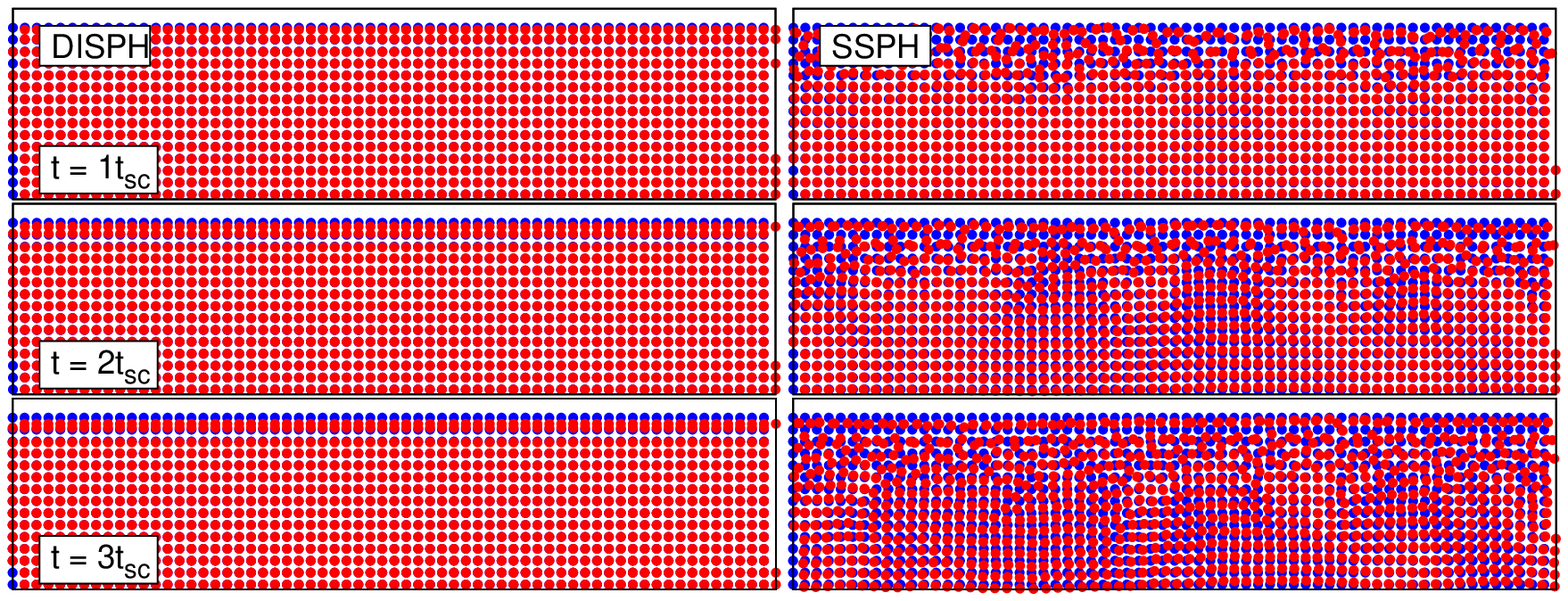}
\caption{
	Snapshots of the free surface tests with DISPH (the left column) and SSPH (the right column).
	The left column shows the results with DISPH, while the right column shows those with SSPH.
	The top, middle and bottom rows show snapshots at $1, 2$ and $3$ sound crossing time.
	The blue circles indicate the identical solution.
	Only particles whose $y$-directional position is greater than $30$ are shown.
}
\label{fig:FS}
\end{figure}

Figure \ref{fig:FS} shows the results of \citet{M94}'s test.
With SSPH, at $t = 1$ sound crossing time, particles around the free surface clearly move to the different positions from the initial positions and at $t = 2-3$ sound crossing time, particles move downward.
With DISPH, similar to SSPH, the outermost particle layer move downward.
However, the other particles virtually keep their initial positions until $t = 3$ and the sound crossing time, second outermost particle layer slightly move upward.
Unlike SSPH, DISPH produces virtually no $x$-directional motions.
It is clear that DISPH can treat the free surface better than SSPH.

\begin{figure}
\plotone{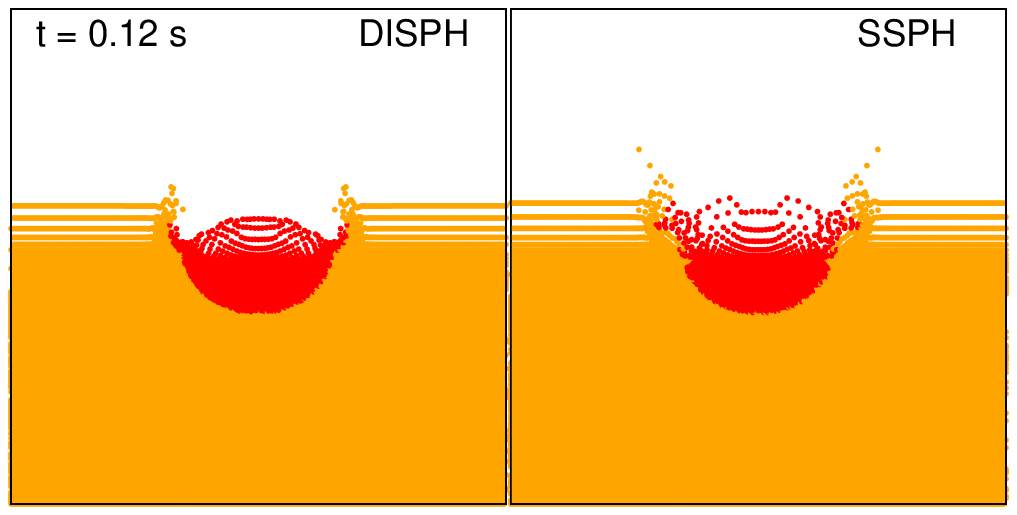}
\caption{
	Snapshots of aluminium-to-aluminium test with DISPH (the left column) and SSPH (the right column) at $t = 0.11$ seconds.
	The orange particles indicate the target particles while the red particles indicate those of impactor particles.
}
\label{fig:Al_to_Al}
\end{figure}

\begin{figure}
\plotone{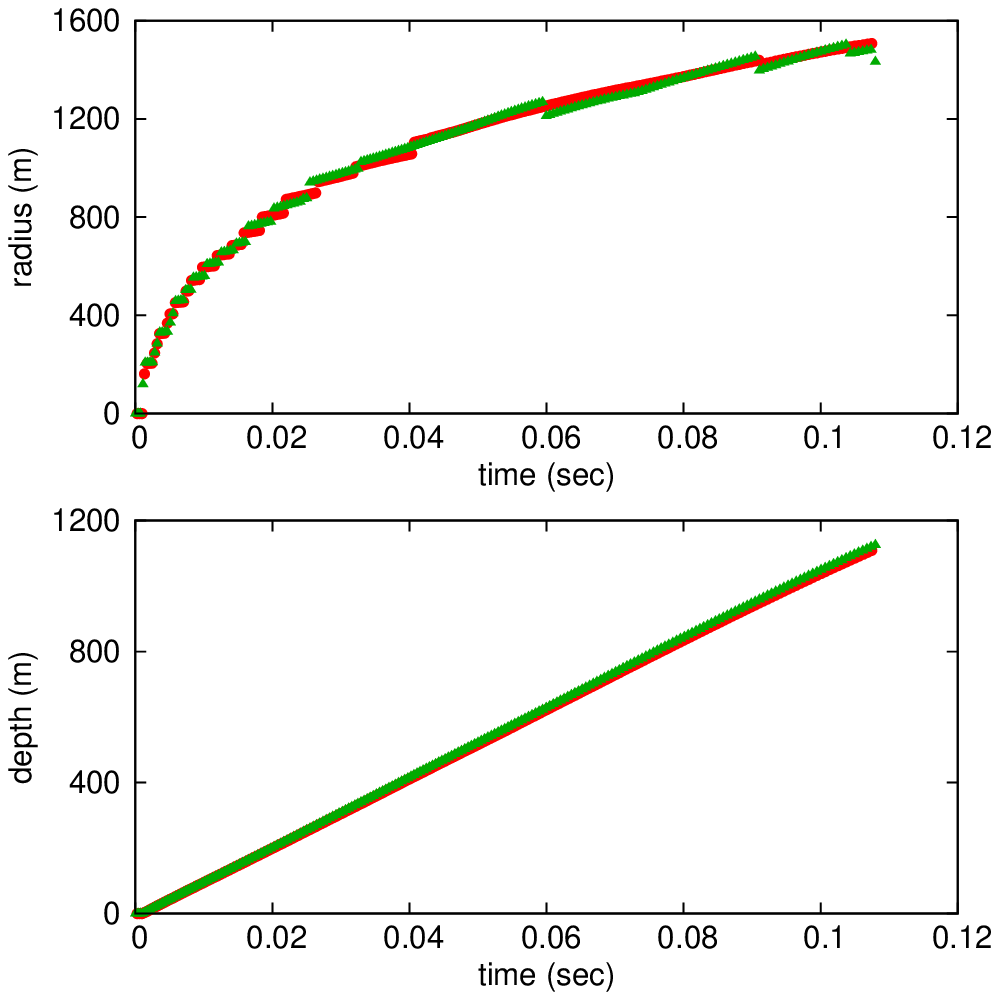}
\caption{
	Time evolutions of crater radius and depth with both methods.
	The red points indicate the results with DISPH, while green points indicate those with SSPH.
}
\label{fig:raddep}
\end{figure}

Figures \ref{fig:Al_to_Al} and \ref{fig:raddep} show the snapshots of the aluminium-to-aluminium test with both methods.
Both methods produce roughly similar results; the jetting and excavation of the target is produced around the impact site.
The crater size and depth are almost indistinguishable between SSPH and DISPH.
DISPH has similar accuracy/errors for free surface as SSPH does.
Note that there are several differences between two results, e.g., the height and expansion of impact jetting.

\begin{figure}
\includegraphics[scale=1.0]{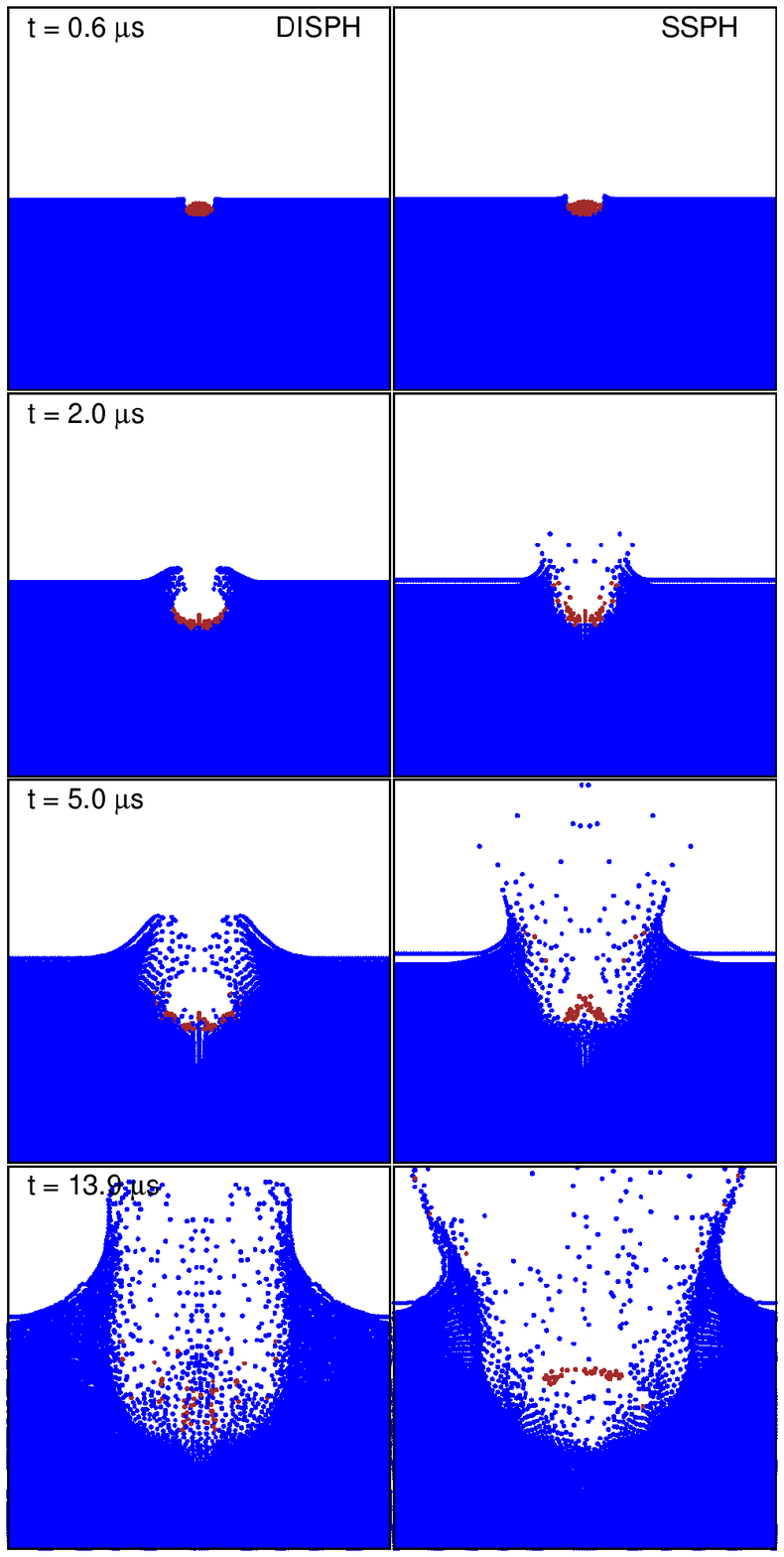}
\caption{
	Snapshots of the glass sphere on water test with DISPH (the left column) and SSPH (the right column).
	The brown particles indicate the projectile particles (wet tuff) while the blue particles indicate those of target particles (water).
}
\label{fig:glass_on_water}
\end{figure}

Figure \ref{fig:glass_on_water} shows the results of the glass-on-water test with both methods.
Unlike the aluminium-to-aluminium test, this test contains the contact discontinuity between water and wet tuff.
Similar to the aluminium-to-aluminium test, the height and expansion of the ejecta curtain is different between two methods, which could be due to the unphysical surface tension between two different materials arising in SSPH calculations.
The target particles are pushed up by the projectile particles at the early step of the impact ($t = 0.6 \mu$s - $2.0 \mu$s).
This results in the higher crater rim with SSPH than DISPH.
At $t = 13.9\mu$s, SSPH produces oblate projectile, while with DISPH, the projectile and target are mixed.
This clearly due to the unphysical surface tension term which results in an underestimate of material mixing.
This difference may be related to the difference in impact-generated disks in the GI simulations between DISPH and SSPH.

\begin{figure}
\plotone{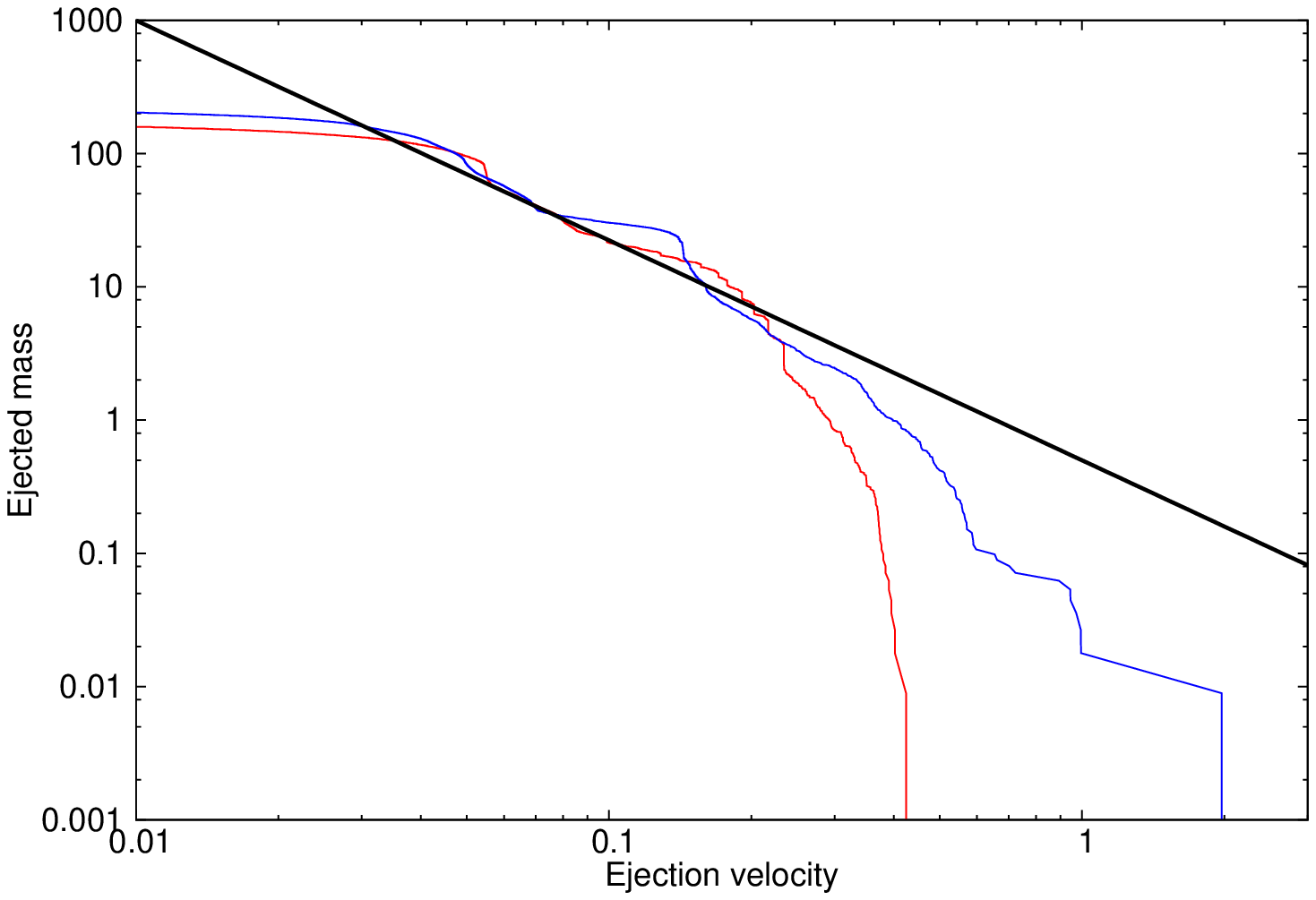}
\caption{
	Mass ejected faster than ejection velocity vs. ejection velocity at $t = 9.0 \mu$s.
	The velocity is normalized by $v^\mathrm{imp} (\rho^\mathrm{imp} / \rho^\mathrm{tar})^{(3\nu - 1)/3 \mu}$ while the mass is normalized by the impactor mass.
	The red curve indicates the result with DISPH, while the blue curve indicates that with SSPH.
	The black line indicates reference line for a power-law with a slope of $v^{- 3 \mu}$.
}
\label{fig:eject_mass}
\end{figure}

Figure \ref{fig:eject_mass} shows the cumulative mass of ejecta $M^\mathrm{ejecta}(v)$ with a vertical velocity greater than a given velocity $v$.
According to the previous works \citep[e.g.,][]{H93, HH11}, the results should have a power-law form with a slope of $3 \mu$:
\begin{eqnarray}
M(v) \propto \left( \frac{v}{v^\mathrm{imp}} \right)^{-3 \mu} \left( \frac{\rho^\mathrm{imp}}{\rho^\mathrm{tar}} \right)^{3\nu - 1},
\end{eqnarray}
where $v^\mathrm{imp}, \rho^\mathrm{imp}$ and $\rho^\mathrm{tar}$ are the velocity of the impactor, the density of the impactor and the density of the target.
Here, $\nu$ and $\mu$ are material parameters which are set to be $0.4$ and $0.55$ \citep{HH11}.
Both methods reproduced roughly similar results to the experiments shown in \citet{HH11}.
The power-law regime with a slope of $-3 \mu$ is well reproduced with both methods.
However, SSPH produces high speed jetting component ($v \gtrsim 0.5$), which can hardly be seen in the experimental results.
This difference should come from the fact that the target particles feel unphysical surface tension from the penetrating projectile, as stated in the previous paragraph.
The target particles are pushed up by the projectile particles to acquire high vertical velocity.
This could result in the difference of the results of GI between two methods.

It is not clear which SPH scheme is more correct, especially for the free surface.
The tests carried out in this Appendix are performed using a 2D cartesian geometry.
The results may differ from 2D cylindrical or 3D geometries.
Thus, it is not straightforward to compare these results with \citet{P+08} and the experiments.
To carry out appropriate comparison, we need to perform 3D impact tests or use 2D axisymmetric domain.
However, note that DISPH does not show unphysical behavior compared to the results with grid code.
We need further investigation to find an appropriate treatment of the free surface, which is left for future work.

\end{document}